\documentclass[aps,prb,twocolumn,superscriptaddress,showpacs]{revtex4}
\usepackage{graphicx}
\usepackage{latexsym}
\usepackage{amssymb}
\usepackage{amsmath}
\usepackage{amsfonts}
\usepackage{bm}
\usepackage{multirow}
\usepackage[usenames,dvipsnames]{color}

\newcommand{\ket}[1]{|#1 \rangle}
\newcommand{\dsZ}{\mathbb{Z}}
\newcommand{\dsR}{\mathbb{R}}
\newcommand{\dsC}{\mathbb{C}}
\newcommand{\dsH}{\mathbb{H}}
\newcommand{\U}{\mathrm{U}(1)}
\newcommand{\dd}{\mathrm{d}}
\newcommand{\ii}{\mathrm{i}}
\newcommand{\Cl}{\mathcal{C\ell}}
\newcommand{\Tr}{\operatorname{Tr}}
\newcommand{\sgn}{\operatorname{sgn}}
\newcommand{\rdim}{\operatorname{rdim}}

\newcommand{\vect}[1]{{\bm{#1}}}
\newcommand{\eqnref}[1]{Eq.\,\eqref{#1}}
\newcommand{\figref}[1]{Fig.\,\ref{#1}}
\newcommand{\tabref}[1]{Tab.\,\ref{#1}}

\begin{document}

\title{Symmetry Protected Topological States of Interacting Fermions and Bosons}

\author{Yi-Zhuang You}

\author{Cenke Xu}

\affiliation{Department of physics, University of California,
Santa Barbara, CA 93106, USA}

\date{\today}

\begin{abstract}

We study the classification for a large class of interacting fermionic and bosonic symmetry protected topological (SPT) states, focusing on the cases where interaction reduces the classification of free fermion SPT states. We define a SPT state as whether or not it is separated from the trivial state through a bulk phase transition, which is a general definition applicable to SPT states with or without spatial symmetries. We show that in all dimensions short range interactions can reduce the classification of free fermion SPT states, and we demonstrate these results by making connection between fermionic and bosonic SPT states. We first demonstrate that our formalism gives the correct classification for several known SPT states, with or without interaction, then we will generalize our method to SPT states that involve the spatial inversion symmetry.

\end{abstract}

\pacs{71.27.+a, 11.10.Kk, 75.40.-s}

\maketitle

\tableofcontents

\section{Introduction}

A symmetry protected topological (SPT) state~\footnote{Sometimes
this kind of states are also called ``symmetry protected trivial"
states in literature, depending on the taste and level of
terminological rigor of authors.} is usually defined as a state
with completely trivial bulk spectrum, but nontrivial ({\it e.g.}
gapless or degenerate) boundary spectrum when and only when the
system including the boundary preserves certain
symmetry~\cite{wenspt1,wenspt2,wenspt3,wenspt4}. The most well-known SPT states
include the Haldane phase of spin-1
chain~\cite{haldane1,haldane2}, quantum spin Hall
insulator~\cite{kane2005a,kane2005b}, topological
insulator~\cite{fukane, moorebalents2007, roy2007}, and
topological superconductor such as Helium-3 B-phase\cite{He3Bbalian,He3Bleggett1,He3Bleggett2,He3Bleggett3}. So far all
the free fermion SPT states have been well understood and
classified in
Ref.~\onlinecite{ludwigclass1,ludwigclass2,kitaevclass,wenclass,morimotoclass}, and
recent studies suggest that interaction 
can reduce the classification of fermionic SPT states~\cite{fidkowski1,fidkowski2,qiz8,zhangz8,levinguz8,yaoz8,chenhe3B,senthilhe3}.
Unlike fermionic systems, bosonic SPT states do need strong
interaction to overcome its tendency to form a Bose-Einstein
condensate. Most bosonic SPT states can be classified by symmetry
group cohomology~\cite{wenspt1, wenspt2, wenspt3,wenspt4}, Chern-Simons
theory~\cite{luashvin,juvengauge} and beyond\cite{juvengauge}, semiclassical non-linear
$\sigma$-model~\cite{xuclass} and cobordism theory\cite{kapustin1,kapustin2}.

The definition for SPT states we gave above is based on the most
obvious phenomenology of the SPT states, and it gives SPT states a
convenient experimental signature, which is their boundary state.
Indeed, the quantum spin Hall insulator and $3d$ topological
insulator were verified experimentally by directly probing their
boundary properties.\cite{KonigQSH, Hsieh:2008qa, Hsieh:2009ix, ChenTI} However, if a SPT state needs certain spatial
symmetry,\cite{fu_inv1,fu_inv2,turner1,fuTCI,taylor,Hsieh:2012sf,turner2,Slager:2013pi,chiuyaoryu} its boundary may be trivial because this spatial
symmetry can be explicitly broken by its boundary. In this work we
will study SPT states both with and without spatial symmetries,
thus in our current work, a SPT state is simply defined as a
gapped and nondegenerate state that must be separated from the
trivial direct product state defined on the same Hilbert space
through one or more bulk phase transitions, as long as the
Hamiltonian always preserves certain symmetry.

In this work we study both strongly interacting fermionic and
bosonic SPT states. In particular, we focus on the instability of  
free fermionic SPT (fFSPT) states against interaction, and investigate their interaction reduced classifications. More exotic fermionic SPT states that can not be reduced from free fermion descriptions\cite{senthilhe3, guwen, xubeyond} are not discussed in this work. The interaction reduced classification of
interacting fermionic SPT (iFSPT) states can be derived by making connection to
bosonic SPT (BSPT) states with the {\it same} symmetry,\endnote{More precisely speaking, we consider the BSPT state with a symmetry action that is compatible with (but not identical to) the symmetry action on the FSPT state. For instance $\dsZ_2^T$ symmetry with $\mathcal{T}^2=-1$ has no bosonic realization, so the corresponding $\dsZ_2^T$ action on BSPT state must be derived from the $\dsZ_2^T$ action on the FSPT state case by case.} and we
will argue that the classification of BSPT states implies the
classification of their fermionic counterparts. More specifically,
since BSPT states always need strong interaction, their
classification tells us how interaction affects the classification
of FSPT states. When describing BSPT states, we will adopt the
formalism developed in Ref.~\onlinecite{xuclass}, namely we
describe a $d$-dimensional BSPT state using an O($d+2$) nonlinear
sigma model (NLSM) field theory with a topological $\Theta$-term,
and we only focus on the stable ``fixed point" states with $\Theta
= 2\pi k$. Depending on integer $k$, these fixed points can
correspond to either trivial or BSPT state. This formalism fits
well with our definition of SPT states: it very naturally tells us
whether two ``fixed point" states can be connected with or without
a phase transition. This is an advantage that we will fully
exploit in our work.

We will first demonstrate our method in section II with well-known
examples such as $1d$ Kitaev's Majorana chain with $\dsZ_2^T$
symmetry~\cite{fidkowski1,fidkowski2}, $2d$ $p\pm\ii p$
topological superconductor (TSC) with $\dsZ_2$
symmetry~\cite{qiz8,yaoz8,levinguz8,zhangz8}, and $3d$ topological
superconductor $^3$He-B with $\dsZ_2^T$
symmetry~\cite{chenhe3B,senthilhe3}. Previous studies show that
although all these states have $\dsZ$ classification without
interaction, their classifications will reduce to $\dsZ_8$,
$\dsZ_8$, $\dsZ_{16}$ under interaction. We will demonstrate that
these interaction-reduced classifications naturally come from the
$\dsZ_2$ classification of $1d$ Haldane spin chain, $2d$ Levin-Gu
paramagnet,\cite{wenspt1, levingu} and $3d$ bosonic SPT state with $\dsZ_2^T$ symmetry.\cite{senthilashvin, xu3dspt}
More precisely, the $\dsZ_2$ classification of the BSPT states
give us a {\it necessary} condition for interaction reduced
classification of their fermionic counterparts. Moreover, the
analysis of BSPT states also leads to a systematic construction of the specific four-fermion
interaction that likely gaps out the bulk critical point between
free fermion SPT (fFSPT) state and the trivial state in the
noninteracting limit. In section III-V we will generalize our
method to SPT states that involve the spatial inversion symmetry. By making connection to BSPT states, we will show that interaction
reduced classification occurs very generally for inversion SPT
states in all dimensions with a systematic pattern.

\section{SPT states without spatial symmetry}

\subsection{From Kitaev Chain to Haldane Chain}

\subsubsection{Lattice Model and Bulk Theory} The Kitaev's Majorana
chain is a $1d$ free fermion SPT (fFSPT) state protected by the
time-reversal symmetry $\dsZ_2^T$ with $\mathcal{T}^2=+1$
(symmetry class BDI)\cite{AZclass}. For generality, we consider
$\nu$ copies of the Majorana chain. The model is defined on a 1d
lattice with $\nu$ flavors of Majorana fermions $\chi_{i\alpha}$
($\alpha=1,\cdots,\nu$) on each site $i$,
\begin{equation}\label{eq: Maj chain real space}
H_{\times\nu}=\sum_{\alpha=1}^\nu\sum_{\langle ij\rangle}\ii
u_{ij} \chi_{i\alpha}\chi_{j\alpha},
\end{equation}
with the bond strength $u_{ij}=u(1+\delta(-)^{i})$ alternating
along the chain. Each unit cell contains two sites, labeled by $A$
and $B$, as shown in \figref{fig: Majorana chain}. The Hamiltonian
is invariant under the time-reversal symmetry $\dsZ_2^T$,
$\mathcal{T}: \chi_{A\alpha}\to\chi_{A\alpha},
\chi_{B\alpha}\to-\chi_{B\alpha}, \ii\to-\ii$, which flips the
sign of the Majorana fermions on the $B$ sublattice followed by
the complexed conjugation.
\begin{figure}[htbp]
\begin{center}
\includegraphics[width=150pt]{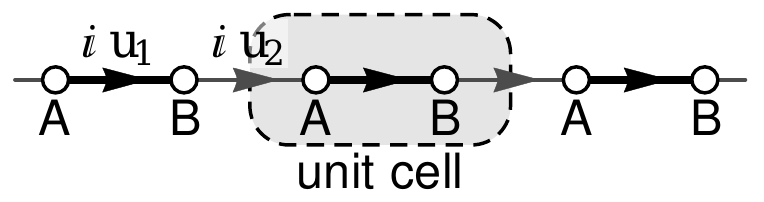}
\caption{Lattice model of Kitaev's Majorana chain.} \label{fig:
Majorana chain}
\end{center}
\end{figure}

Fourier transform to the momentum space and introduce the basis
$\chi_{k\alpha}=(\chi_{kA\alpha},\chi_{kB\alpha})^\intercal$, the
Hamiltonian \eqnref{eq: Maj chain real space} becomes
\begin{equation}\label{eq: Maj chain k space}
H_{\times
\nu}=\frac{1}{2}\sum_{\alpha=1}^{\nu}\sum_{k}\chi_{-k\alpha}^{\intercal}\left[\begin{matrix}0
& -\ii u_{k}^*\\ \ii u_{k} & 0\end{matrix}\right]\chi_{k\alpha},
\end{equation}
with $\ii u_{k}=\ii u(1+\delta)-\ii u(1-\delta)e^{-\ii k}$, and
the time-reversal symmetry acts as
$\mathcal{T}:\chi_{k\alpha}\to\sigma^3\chi_{-k\alpha}$. In this
paper, we use $\sigma^1$, $\sigma^2$, $\sigma^3$ to denote the
Pauli matrices. When $\delta \ll 1$, in the long-wave-length limit
($k\to0$), $\ii u_{k}\to u(-k+2\ii\delta)$, so the low-energy
effective Hamiltonian reads
\begin{equation}\label{eq: 1d H fFSPT}
H_{\times\nu}=\frac{1}{2}\sum_{\alpha=1}^{\nu}\int\dd
x\;\chi_{\alpha}^\intercal(\ii\partial_x\sigma^1+m\sigma^2)\chi_{\alpha}.
\end{equation}
Here we have set $u=1$ and introduced $m=2\delta$ as the
topological mass. The time-reversal symmetry operator may be
written as $\mathcal{T}=\mathcal{K}\sigma^3$, where
$\mathcal{K}^{-1}\ii\mathcal{K}=-\ii$ implements the complex
conjugation by flipping the imaginary unit.

On the free fermion level, the $1d$ BDI class FSPT states are
$\dsZ$ classified,\cite{ludwigclass1,ludwigclass2} as indexed by a
bulk topological integer
\begin{equation}
N=\int \frac{\dd k}{4\pi\ii}\;\Tr\sigma^3
G(k)\partial_{k}G(k)^{-1},
\end{equation}
where $G(k)=-\langle\chi_{k}\chi_{-k}^\intercal\rangle$ is the
fermion Green's function at zero frequency. Given the model in
\eqnref{eq: Maj chain k space}, the topological number $N=\nu
(1-\sgn \delta)/2$ is identical to the fermion flavor number $\nu$ when
$\delta<0$. Every topological number $N\in\dsZ$ indexes a distinct
fFSPT phase, which, in respect of the symmetry, can not be
connected to each other without going through a bulk transition.

\subsubsection{Corresponding BSPT state}

However with interaction, the classification can be reduced from
$\dsZ$ to $\dsZ_8$, meaning that eight copies of the Majorana
chain ($N=8$) can be smoothly connect to the trivial state ($N=0$)
without closing the bulk gap in the presence of interaction. This
interaction reduced classification was discovered in
Ref.\,\onlinecite{fidkowski1,fidkowski2}. But we will show it
again by making connection to the boson SPT (BSPT) states, as our
approach can be generalized to higher spacial dimensions.

Let us start by showing that four copies of the Majorana chain
($\nu=4$) can be connected to the Haldane spin chain,\cite{xubf} a
$\dsZ_2^T$ BSPT state in 1d. In this case, we have four flavors of
Majorana fermion per site, denoted $\chi_{i\alpha}$
($\alpha=1,\cdots,4$), from which we can define a spin-1/2 object
on each site, as
$\vect{S}_{i}=(S_{i1},S_{i2},S_{i3})=\chi_{i}^{\intercal}(\sigma^{12},\sigma^{20},\sigma^{32})\chi_{i}$
in the basis
$\chi_{i}=(\chi_{i1},\chi_{i2},\chi_{i3},\chi_{i4})^\intercal$.
\footnote{Through out this paper, we use the notation
$\sigma^{ijk\cdots}\equiv\sigma^i\otimes\sigma^j\otimes\sigma^k\otimes\cdots$
for the Kronecker product (direct product) of the Pauli matrices,
where $\sigma^1$, $\sigma^2$, $\sigma^3$ stands for the three
Pauli matrices respectively while $\sigma^0$ denotes the
$2\times2$ identity matrix.} One can couple the staggered
component of the spin to an O(3) order parameter $\vect{n}_i$ on
each site, as $H_{BF}=-\sum_{i} (-)^i\vect{n}_i\cdot\vect{S}_i$.
Then the low-energy effective Hamiltonian for four copies of the
Majorana chain coupled to the $\vect{n}$ field reads (which we
call fermionic $\sigma$-model, or FSM)
\begin{equation}\label{eq: 1d Maj FSM}
\begin{split}
H_{\times4}&=\frac{1}{2}\int\dd x\;\chi^\intercal h_{\times4}\chi,\\
h_{\times4}&=\ii\partial_1\sigma^{100}+m\sigma^{200}+n_1\sigma^{312}+n_2\sigma^{320}+n_3\sigma^{332},
\end{split}
\end{equation}
where $\chi=(\chi_{A},\chi_{B})^\intercal$. The time-reversal
symmetry operator $\mathcal{T}=\mathcal{K}\sigma^{300}$
necessarily requires to flip the order parameters
$\vect{n}\to-\vect{n}$ under the time-reversal. Following the
calculation in Ref.~\onlinecite{abanov2000}, after integrating out
the fermion field $\chi$, we arrive at the effective theory for
the boson field $\vect{n}$, which is a non-linear $\sigma$-model
(NLSM) with a topological $\Theta$ term at $\Theta=2\pi$, given by
the following action
\begin{equation}
S[\vect{n}]=\int \dd\tau\dd x\;\frac{1}{g}(\partial_\mu
\vect{n})^2+\frac{\ii\Theta}{4\pi}\epsilon^{abc}n_a\partial_{\tau}n_b\partial_{x}n_c.
\label{1do3}
\end{equation}
$g^{-1}(\partial_\nu\vect{n})^2$ describes the remaining dynamics
in the bosonic sector. Presumably we work in the large
$g\to\infty$ limit, such that the $\vect{n}$ field is deep in its
disordered phase. With $\Theta = 0$, the Hamiltonian of
\eqnref{1do3} reads $H = \int dx \ g \vect{L}^2 + \frac{1}{g}
(\nabla_x \vect{n})^2$ (where $\vect{L}(x)$ is the canonical
conjugate variable of $\vect{n}(x)$ at each spatial position $x$),
and since $g$ flows to $+\infty$ under coarse-graining, in the
long-wave-length limit the ground state wave function of this
theory is a trivial direct product state $|\Omega\rangle = \prod_x
|l = 0\rangle$~\cite{xu3dspt} with a fully gapped and
nondegenerate spectrum in the bulk and at the boundary (on each
coarse grained spatial point, $\vect{L}^2 = l(l+1)$). However,
with $\Theta = 2\pi$, \eqnref{1do3} describes a non-trivial BSPT
state for the $\vect{n}$ field, and is
equivalent~\cite{haldane1,haldane2,ng1994} to the Haldane phase of
spin-1 chain protected by the spin-flipping time-reversal symmetry
$\mathcal{T}: \vect{n}\to-\vect{n}$. In fact, the spatial boundary
of \eqnref{1do3} with $\Theta = 2\pi$ is a $(0+1)d$ O(3) NLSM with
a Wess-Zumino-Witten (WZW) term at level $k=1$, and by solving
this theory exactly, we can demonstrate explicitly that the ground
state of the spatial boundary of \eqnref{1do3} with $\Theta =
2\pi$ is doubly degenerate~\cite{ng1994,xuclass}, which is
equivalent to the boundary ground state of four copies of Kitaev's
chain under interaction. In the low-energy limit, the boundary of
four copies of the FSPT states is faithfully captured by the
bosonic field $\vect{n}$. Thus we have established a connection
between four copies of Majorana chain and a single copy of Haldane
chain, bridging the FSPT and BSPT states in $1d$.

Using the knowledge of the better understood BSPT states, we can
gain insight of the interacting fermion SPT (iFSPT) states. If
eight copies of the iFSPT states is a trivial phase, then {\it
necessarily} the bosonic theory of eight copies of the FSPT states
derived using the same method must also be a trivial state.
Indeed, because the Haldane phase has a well-known $\dsZ_2$
classification~\cite{wenspt2}, it is expected that two copies of
the Haldane chain can be smoothly connected to the trivial state
without breaking the symmetry. This can be shown by coupling two
layers of the Haldane chain with a large inter-layer
anti-ferromagnetic interaction (which preserves the $\dsZ_2^T$
symmetry), as described by the action
\begin{equation}
\begin{split}
S&=S[\vect{n}^{(1)}]+S[\vect{n}^{(2)}]+S_\text{cp},\\
S_\text{cp}&=\int \dd\tau\dd
x\;A\vect{n}^{(1)}\cdot\vect{n}^{(2)},
\end{split} \label{1do3c}
\end{equation}
when $A \to+\infty$, $\vect{n}^{(1)}$ and $\vect{n}^{(2)}$ are
locked into opposite directions, \emph{i.e.} $\vect{n}^{(1)} =
-\vect{n}^{(2)} = \vect{n}$. Then the effective NLSM for
$\vect{n}$ has $\Theta = 0$ due to the cancellation of the
$\Theta$ angles between the two layers. So two copies of the
Haldane chain can be trivialized by the $A$ coupling.

Also, when the two Haldane phases in \eqnref{1do3c} are decoupled
from each other ($A = 0$), both Haldane phases in \eqnref{1do3c}
are separated from the trivial phase ($\Theta = 0$) with a
critical point at $\Theta = \pi$. However, with $A \neq 0$, this
critical point is also gapped out by the $A$
coupling,\footnote{When $\Theta = \pi$, both $S[\vect{n}^{(1)}]$
and $S[\vect{n}^{(2)}]$ can be viewed as the low energy field
theory of spin-1/2 chains. Then the antiferromagnetic inter-chain
coupling $A$ will drive the system into a fully gapped state which
is a direct product of inter-chain spin singlet on every site.}
thus with $A
> 0$, the entire phase diagram of \eqnref{1do3c} has only one
trivial phase. This observation already suggests that 8 copies of
Kitaev's Majorana chain is trivial under interaction.


\subsubsection{Bulk transition}

Now let us carefully investigate the interactions in the fermion
model. Two copies of the Haldane chain would correspond to eight
copies of the Majorana chain. Recall the relation
$\vect{n}_{i}\sim(-)^i\langle\vect{S}_i\rangle$ on the mean-field
level, the inter-layer coupling $A$ can be immediately ported to
the fermion model as an on-site interaction among eight flavors of
Majorana fermions
\begin{equation}\label{eq: Hint 1d}
\begin{split}
&H_\text{int}=\frac{J}{4}\sum_{i}\vect{S}_{i}^{(1)}\cdot\vect{S}_{i}^{(2)}\\
&=J\sum_{i}( -\chi_{i1}
\chi_{i2}\chi_{i5}\chi_{i6}+\chi_{i1}\chi_{i2}\chi_{i7}\chi_{i8}-\chi_{i1}\chi_{i3}\chi_{i5}\chi_{i7}\\&-\chi_{i1}\chi_{i3}\chi_{i6}\chi_{i8}-\chi_{i1}\chi_{i4}\chi_{i5}\chi_{i8}+\chi_{i1}\chi_{i4}\chi_{i6}\chi_{i7}\\&+\chi_{i2}\chi_{i3}\chi_{i5}\chi_{i8}-\chi_{i2}\chi_{i3}\chi_{i6}\chi_{i7}-\chi_{i2}\chi_{i4}\chi_{i5}\chi_{i7}\\&-\chi_{i2}\chi_{i4}\chi_{i6}\chi_{i8}+\chi_{i3}\chi_{i4}\chi_{i5}\chi_{i6}-\chi_{i3}\chi_{i4}\chi_{i7}\chi_{i8}),
\end{split}
\end{equation}
with $J>0$. We should expect that eight copies of the Majorana
chain can be connected to the trivial state under this
interaction, as the same interaction can trivialize the BSPT in
the NLSM.

\begin{figure}[htbp]
\begin{center}
\includegraphics[width=200pt]{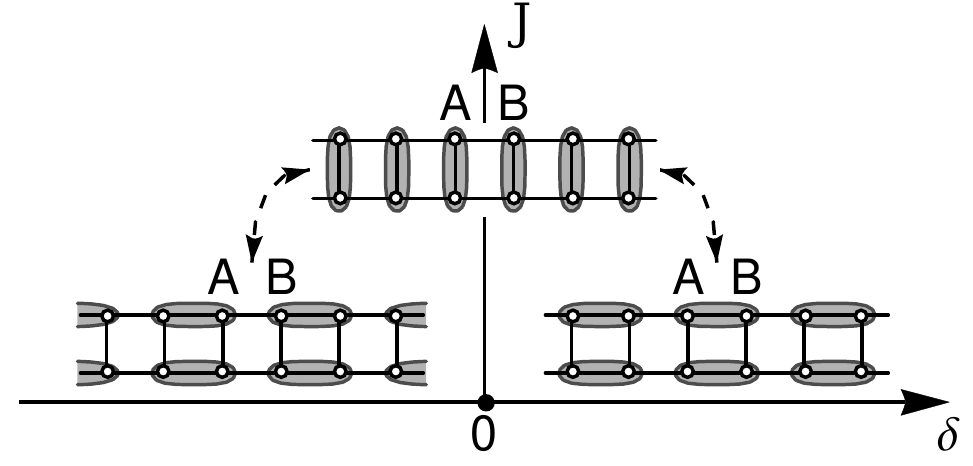}
\caption{Phase diagram of the interacting Majorana chain at
$\nu=8$. The bulk criticality at the origin can be avoided if
interaction is allowed. Each horizontal chain is four copies of
the Majorana chain, equivalent to a Haldane chain. The vertical
bound is the on-site spin-spin interaction. $A$ and $B$ labels the
sites in a unit cell. Gray ovals mark out the spin-singlet
dimers.} \label{fig: dimer}
\end{center}
\end{figure}

Such an expectation is obvious in the spin sector. In the free
fermion limit, depending on the sign of $\delta$, the $\nu=8$
fFSPT states may correspond to two different spin-singlet
dimerization patterns: the intra-unit-cell dimerization
($\delta>0$ trivial state) or the inter-unit-cell dimerization
($\delta<0$ SPT state), as shown in \figref{fig: dimer}. While the
strong on-site interaction in \eqnref{eq: Hint 1d} will lead to a
third pattern, \emph{i.e.} the on-site (inter-layer) dimerization,
see \figref{fig: dimer}. The three patterns are connected by the
ring exchange of the dimmers. However, it is known that the ring
exchange is a smooth deformation and will not close the spin gap,
so at least in the spin sector, the $\delta<0$ and the $\delta>0$
SPT states can be smoothly connected.

\begin{figure}[htbp]
\begin{center}
\includegraphics[width=150pt]{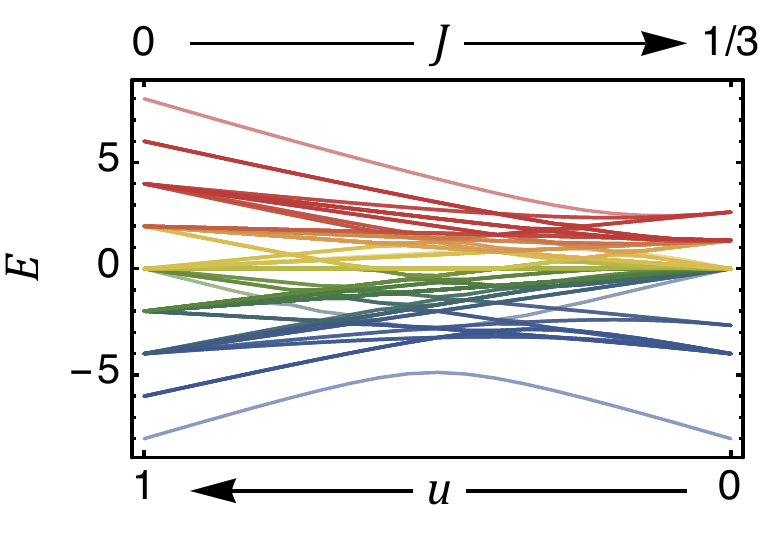}
\caption{Many-body energy levels of the two-site segment by exact
diagonalization. From the weak interaction limit (left) to the
strong interaction (right) limit, the many-body gap never closes.}
\label{fig: two site}
\end{center}
\end{figure}

To show that the charge gap also remains open, we can perform an
explicit calculation based on the lattice model \eqnref{eq: Maj
chain real space} in the strong dimerization limit $\delta=\pm1$,
such that the 1d chain is decoupled into independent two-site
segments. In each segment, the interacting fermion system can be
exact diagonalized. Then it can be shown that the charge gap
indeed persists as $u$ is tuned to zero in the present of the
interaction $J$, as shown in \figref{fig: two site}. So one can
smoothly connect the $N=8$ fFSPT state to the $N=0$ fFSPT state in
three steps: (i) turn on $J$ and turn off $u$, (ii) change the
sign of $\delta$, (iii) turn on $u$ and turn off $J$. The bulk gap
will never close during this process. Thus the whole phase diagram
in \figref{fig: dimer} is actually one phase.

In conclusion, we have demonstrated that the classification of the
$1d$ FSPT states with the (BDI class) time-reversal symmetry is
reduced from $\dsZ$ to $\dsZ_8$ under interation. We obtain the
iFSPT classification by making connection to the BSPT
classification. This approach can be readily generalized to higher
spacial dimensions in the following. Moreover, the way that the
BSPT state can be trivialized in the NLSM naturally provide us the
correct fermion interaction that is needed to trivialize the FSPT
states, which can be much more general than the currently known
Fidkowski-Kitaev type of interaction. And this interaction can gap
out the critical point $m = 0$ in \eqnref{eq: 1d H fFSPT}, which
is 8 copies of nonchiral $1d$ Majorana fermions. This bulk
analysis is particularly suitable to study the crystalline SPT
states, which may not have symmetry protected physical boundary
modes.

\subsection{From $2d$ TSC to Levin-Gu Paramagnet}

\subsubsection{Lattice Model and Bulk Theory}

Now we turn to the $2d$ example of the $p\pm\ii p$ topological
superconductor (TSC) protected by a $\dsZ_2$ symmetry (symmetric
class D)\cite{AZclass}. The $p_x\pm\ii p_y$ TSC can be viewed as
both layers of the $p_x+\ii p_y$ and the $p_x-\ii p_y$
TSC's\cite{volovik,ReadGreen,fukanep+ip} stacked together with the
$\dsZ_2$ symmetry acts only in the $p_x-\ii p_y$ layer by flipping
the sign of the fermion operator (\emph{i.e.} the fermion parity
transform). On a $2d$ lattice, the model Hamiltonian can be
written as
\begin{equation}\label{eq: 2d TSC}
\begin{split}
H&=\sum_{\vect{k},l}\xi_\vect{k}c_{\vect{k}l}^\dagger c_{\vect{k}l}+\frac{1}{2}(\Delta_\vect{k}c_{-\vect{k}l}c_{\vect{k}l}+h.c.),\\
\xi_\vect{k}&=-2t(\cos k_1+\cos k_2)-\mu,\\
\Delta_{\vect{k}l}&=-\Delta (\sin k_1+(-)^l\ii \sin k_2),
\end{split}
\end{equation}
where $l=0,1$ labels the two opposite layers of the chiral TSC's.
The $\dsZ_2$ symmetry acts as $c_{\vect{k}l}\to(-)^lc_{\vect{k}l}$
which prevents the mixing of fermions from different layers, such
that the fermion parity is conserved in each layer independently.
Depending on the chemical potential $\mu$, the model has two
phases: the $|\mu|>4t$ strong pairing trivial superconductor phase
and the $|\mu|<4t$ weak pairing topological superconductor phase.
Switching to the Majorana basis
$\chi_{\vect{k}}=(c_{\vect{k}0}+c_{-\vect{k}0}^\dagger,
c_{\vect{k}1}+c_{-\vect{k}1}^\dagger, \ii c_{\vect{k}0}-\ii
c_{-\vect{k}0}^\dagger, \ii c_{\vect{k}1}-\ii
c_{-\vect{k}1}^\dagger)^\intercal/\sqrt{2}$ and in the
long-wave-length limit, the effective Hamiltonian reads
\begin{equation}
H_{\times1}=\frac{1}{2}\sum_{\vect{k}}\chi_{-\vect{k}}^\intercal(-k_1\sigma^{30}-k_2\sigma^{13}+m\sigma^{20})\chi_{\vect{k}},
\end{equation}
where we have set $\Delta=1$ as our energy unit, and defined the
topological mass $m=-4t-\mu$ (assuming $\mu<0$). The $\dsZ_2$
symmetry acting on the Majorana fermions as
$\chi_{\vect{k}}\to\sigma^{03}\chi_{\vect{k}}$. The trivial
($m>0$) and the topological ($m<0$) phases are separated by the
phase transition at $m=0$ where the bulk gap closes. This bulk
criticality is protected by the $\dsZ_2$ symmetry.

The above $p\pm\ii p$ TSC is an example of the $2d$ D class fFSPT
states, which are known to be $\dsZ$
classified,\cite{ludwigclass1,ludwigclass2} and are indexed by the
topological number\cite{qi2006}
\begin{equation}
N=\int\frac{\dd^3k}{8\pi^2}\Tr\sigma^{03}G\partial_{\ii\omega}G^{-1}G\partial_{k_1}G^{-1}G\partial_{k_2}G^{-1},
\end{equation}
where $G(k)=-\langle\chi_{k}\chi_{-k}^\intercal\rangle$ with
$k=(\ii\omega,\vect{k})$ is the fermion Green's function in the
frequency-momentum space. The $p\pm\ii p$ TSC in \eqnref{eq: 2d
TSC} corresponds to $N=1$. While the other topological states in
this $\dsZ$ classification may be realized by considering multiple
copies of such $p\pm\ii p$ TSC's, which can be described by the
following effective field theory Hamiltonian
\begin{equation}\label{eq: 2d H fFSPT}
H_{\times\nu}=\frac{1}{2}\sum_{\alpha=1}^{\nu}\int\dd^2
\vect{x}\;\chi_\alpha^\intercal(\ii\partial_1\sigma^{30}+\ii\partial_2\sigma^{13}+m\sigma^{20})\chi_\alpha,
\end{equation}
with the $\dsZ_2: \chi_\alpha\to\sigma^{03}\chi_\alpha$ symmetry
protection. $\nu$-copy $p\pm\ii p$ TSC would correspond to the
topological number $N=\nu$.

\subsubsection{Corresponding BSPT state}

However with interaction, the classification of the $\dsZ_2$
$p\pm\ii p$ TSC is reduced from $\dsZ$ to
$\dsZ_8$~\cite{qiz8,zhangz8,yaoz8,levinguz8}, meaning that eight
copies of the $p\pm\ii p$ TSC ($N=8$) can be smoothly connected to
the trivial state ($N=0$) in the presence of interaction. This
interaction reduced classification was discussed in
Ref.\,\onlinecite{qiz8,yaoz8,zhangz8,levinguz8}, but here we will
provide another argument for it by making connection to $2d$ BSPT
states.

Let us start by showing that four copies of the $p\pm\ii p$ TSC
($\nu=4$) can be connected\cite{luashvin,chenggu} to the Levin-Gu topological
paramagnet,\cite{wenspt1, levingu} a $\dsZ_2$ BSPT state in $2d$. We first
introduce a set of inter-layer $s$-wave pairing terms (with
$\bar{l}\equiv1-l$ and $\alpha,\alpha'=1,2,3,4$ labeling the 4
copies)
\begin{equation}\label{eq: O4 orders}
\begin{split}
\Delta_1&=\sum_{l,\alpha,\alpha'}c_{\bar{l}\alpha}\ii\sigma^{12}_{\alpha\alpha'}c_{l\alpha'}+h.c.=\chi^\intercal\sigma^{1112}\chi,\\
\Delta_2&=\sum_{l,\alpha,\alpha'}c_{\bar{l}\alpha}\ii\sigma^{20}_{\alpha\alpha'}c_{l\alpha'}+h.c.=\chi^\intercal\sigma^{1120}\chi,\\
\Delta_3&=\sum_{l,\alpha,\alpha'}c_{\bar{l}\alpha}\ii\sigma^{32}_{\alpha\alpha'}c_{l\alpha'}+h.c.=\chi^\intercal\sigma^{1132}\chi,\\
\Delta_4&=\sum_{l,\alpha} c_{\bar{l}\alpha}(-)^l
c_{l\alpha}+h.c.=\chi^\intercal\sigma^{1200}\chi,
\end{split}
\end{equation}
where $\chi=(\chi_1,\chi_2,\chi_3,\chi_4)^\intercal$; and couple
them to an O(4) order parameter field
$\vect{n}=(n_1,n_2,n_3,n_4)$. The low-energy effective Hamiltonian
of this FSM reads
\begin{equation}\label{eq: 2d TSC FSM}
\begin{split}
H_{\times4}&=\frac{1}{2}\int\dd^2 \vect{x}\;\chi^\intercal h_{\times4}\chi,\\
h_{\times4}&=\ii\partial_1\sigma^{3000}+\ii\partial_2\sigma^{1300}+m\sigma^{2000}\\
&+n_1\sigma^{1112}+n_2\sigma^{1120}+n_3\sigma^{1132}+n_4\sigma^{1200}.
\end{split}
\end{equation}
Because the inter-layer pairing mixes the fermions between the
$p_x+\ii p_y$ and the $p_x-\ii p_y$ TSC's, they will gain a minus
sign under the $\dsZ_2$ symmetry transform. To preserve the
$\dsZ_2$ symmetry, we must require the order parameters to change
sign as well, \emph{i.e.} $\vect{n}\to-\vect{n}$, under the
$\dsZ_2$ symmetry action. After integrating out the fermion field
$\chi$, we arrive at the effective theory for the boson field
$\vect{n}$, which is a NLSM with a topological $\Theta$ term at
$\Theta=2\pi$, given by the following action
($\dd^3x=\dd\tau\dd^2\vect{x}$)
\begin{equation}\label{eq: 2d NLSM}
S[\vect{n}]=\int \dd^3 x\;\frac{1}{g}(\partial_\mu
\vect{n})^2+\frac{\ii\Theta}{2\pi^2}\epsilon^{abcd}n_a\partial_0
n_b\partial_1n_c\partial_2n_d,
\end{equation}
which describes a non-trivial BSPT state for the $\vect{n}$
field~\cite{xusenthil,xuclass}, and is equivalent to the Levin-Gu
state\cite{xuclass} protected by the $\dsZ_2$ symmetry
$\vect{n}\to-\vect{n}$. This can be understood from the wave
function perspective. We first reparameterize
$\vect{n}=(\vect{m}\cos \alpha, \phi \sin\alpha)$ where
$\vect{m}=(m_1,m_2,m_3)$ is an O(3) unit vector and $\phi=\pm1$.
Suppose the system energetically favors $\vect{m}$ (\emph{i.e.}
$\alpha=0$), then the wave function for the $\vect{m}$ field in
its paramagnetic phase ($g\to\infty$) can be derived from the
action \eqnref{eq: 2d NLSM} as\cite{xusenthil}
\begin{equation}
\begin{split}
\ket{\Psi}\sim&\int\mathcal{D}[\vect{m}] e^{\int\dd^2 \vect{x}\frac{\ii \pi}{4\pi}\epsilon^{abc}m_a\partial_1m_b\partial_2m_c}\ket{[\vect{m}]}\\
=&\int\mathcal{D}[\vect{m}] (-)^{N_s[\vect{m}]}\ket{[\vect{m}]}\\
\sim&\int\mathcal{D}[m_3] (-)^{N_d[m_3]}\ket{[m_3]},
\end{split}
\end{equation}
which is a superposition of all $\vect{m}$ configurations with a
sign factor $(-)^{N_s[\vect{m}]}$ counting the parity of the
Skyrmion number $N_s$ of the $\vect{m}$ field. In the Ising limit
where $m_3$ is energetically favored, the Skyrmion number $N_s$
becomes the domain-wall number $N_d$ of the Ising spin $m_3$, so
the wave function becomes the superposition of Ising
configurations with the domain-wall sign~\cite{xusenthil}, which
is exactly the wave function of the Levin-Gu state~\cite{levingu}.
Thus we have established a connection from four copies of the
$p\pm\ii p$ TSC to a single copy of the Levin-Gu paramagnet,
bridging the FSPT and BSPT states in $2d$.

Now we can discuss the iFSPT states using the knowledge about the
BSPT states: if eight copies of the TSC is trivial, then the
bosonic theory derived using the same method above must {\it
necessarily} be trivial. Indeed, on the BSPT side, we know that
two copies of the Levin-Gu paramagnets can be smoothly connected
to the trivial state without breaking the symmetry, which can be
realized by coupling two layers of the Levin-Gu paramagnet with a
large inter-layer anti-ferromagnetic interaction, such that the
domain-wall configuration in both layers will become identical,
and the domain-wall sign from both layers will cancel out, so that
the resulting wave function is just a trivial Ising paramagnetic
state. At the field theory level, it can be described by the
following action with inter-layer coupling
\begin{equation}
\begin{split}
S=&S[\vect{n}^{(1)}]+S[\vect{n}^{(2)}]+S_\text{cp},\\
S_\text{cp}=&\int \dd^3 x\;A(n_1^{(1)}n_1^{(2)}+n_2^{(1)}n_2^{(2)}+n_3^{(1)}n_3^{(2)})\\
&\phantom{\int \dd^3 x\;}-Bn_4^{(1)}n_4^{(2)}.
\end{split}
\end{equation}
It is easy to check that the coupling $S_\text{cp}$ respects the
$\dsZ_2$ symmetry. When $A,B\to+\infty$, $\vect{n}^{(1)}$ and
$\vect{n}^{(2)}$ are locked anti-ferromagnetically for their first
three components and ferromagnetically for their last components,
\emph{i.e.} $n_{a}^{(1)} = -n_{a}^{(2)} = n_a$ ($a=1,2,3$) and
$n_{4}^{(1)} = n_{4}^{(2)} = n_4$. Then the effective NLSM for the
combined field $\vect{n}$ has $\Theta = 0$ due to the cancellation
of the $\Theta$ angles between the two layers. So two copies of
the Levin-Gu paramagnet can be trivialized by the $A, B\to
+\infty$ coupling~\footnote{As one can see, the design of the
coupling is not unique, any inter-layer coupling that locks odd
number of $\vect{n}$ components anti-ferromagnetically will do the
job to trivialize the BSPT state (for example $A,B\to-\infty$ is
also a choice), but here let us stick to our current design and
focus on the $A, B\to +\infty$ coupling.}. This suggests that
eight copies of the original $p\pm ip$ TSC is trivial.

\subsubsection{Boundary modes and Bulk transition}

Recall the relation $n_a\sim\langle\Delta_a\rangle$ ($a=1,2,3,4$)
on the mean-field level, the inter-layer coupling $S_\text{cp}$
can be immediately ported to the fermion model as the following
four-fermion interaction (with $A,B>0$)
\begin{equation}\label{eq: Hint 2d}
H_\text{int}=\int\dd^2\vect{x}\;A\sum_{a=1,2,3}\Delta_{a}^{(1)}\Delta_{a}^{(2)}-B\Delta_{4}^{(1)}\Delta_{4}^{(2)},
\end{equation}
where $\Delta_{a}$ is defined in \eqnref{eq: O4 orders}.
Without any interaction, eight copies of $p \pm ip$ TSC with the
$\dsZ_2$ symmetry is separated from the trivial state through a
critical point that has 16 copies of $2d$ massless Majorana
fermions in the bulk ($m = 0$ in \eqnref{eq: 2d H fFSPT}). We
should expect that the bulk criticality can be gapped out by the
interaction \eqnref{eq: Hint 2d}, and eight copies of the $p\pm\ii
p$ TSC can be smoothly connected to the trivial state, as the same
interaction can trivialize the BSPT in the NLSM.

Admittedly, in $2d$ (and higher dimensions), it is hard to
explicitly demonstrate how the interaction gaps out the gapless
bulk fermion at the critical point. Nevertheless we can show that,
on an open manifold, the interaction Eq.~\eqref{eq: Hint 2d} can
gap out the $1d$ boundary states of eight copies of the $p\pm\ii
p$ TSC ($N=8$) without breaking the symmetry, and hence there
should be no obstacle to tune the bulk system smoothly from the
$N=8$ state to the $N=0$ state under interaction. The
``transition" between $N = 8$ and $N = 0$ states can be viewed as
growing $N = 0$ domains inside the $N = 8$ state, which is
equivalent to sweeping the interface between the two states
through the entire bulk (this is essentially the picture of
Chalker-Coddington model~\cite{cc1988} for the quantum Hall
plateau transition), then as long as the interface is gapped out
by interaction, the bulk gap never has to close during this
``transition", namely the bulk phase transition can be gapped out
by the interaction. Thus all we need to show here is that the
interaction \eqnref{eq: Hint 2d} induces an effective interaction
at the $1d$ boundary, which will gap out the boundary states.

Let us consider a boundary of the $2d$ system along the $x_2$
axis, \emph{i.e.} the topological mass $m\sim x_1$ changes sign
across $x_1=0$. For four copies of the $p\pm\ii p$ TSC as
described in \eqnref{eq: 2d TSC FSM}, the boundary states are
given by the projection operator
$P=(1-\sigma^{3000}\sigma^{2000})/2$, such that the effective FSM
Hamiltonian along the boundary is given by
\begin{equation}
\begin{split}
H'_{\times4}&=\frac{1}{2}\int\dd x_2 \eta^\intercal h'_{\times 4}\eta,\\
h'_{\times4}&=\ii\partial_2\sigma^{300}+n_1\sigma^{112}+n_2\sigma^{120}+n_3\sigma^{132}+n_4\sigma^{200},
\end{split}
\end{equation}
where $\eta$ denotes the Majorana edge modes, and the $\dsZ_2$
symmetry acts as $\eta\to\sigma^{300}\eta$. Under a basis
transformation $\eta\to \exp(-\frac{\ii\pi}{4}\sigma^{200})\eta$,
the boundary FSM Hamiltonian can be reformulated as
\begin{equation}\label{eq: 2d TSCx4 boundary}
h'_{\times4}=-\ii\partial_2\sigma^{100}+n_1\sigma^{312}+n_2\sigma^{320}+n_3\sigma^{332}+n_4\sigma^{200},
\end{equation}
which, at the field theory level, is equivalent to four copies of
$1d$ (critical) Majorana chain described by \eqnref{eq: 1d Maj
FSM}, with the transformed $\dsZ_2$ symmetry
$\eta\to-\sigma^{100}\eta$. $(n_1,n_2,n_3)$ is the analogue of the
O(3) order parameter of the Majorana chain introduced in the
previous section. All these order parameters are forbidden to
condense by the $\dsZ_2$ symmetry, \emph{i.e.}
$\langle\vect{n}\rangle=0$, so that the edge is gapless at the
free fermion level.


Now we consider the boundary of eight copies of the $p\pm\ii p$
TSC, which is simply a doubling of \eqnref{eq: 2d TSCx4 boundary}.
The field theory of this $1d$ boundary is equivalent to eight
copies of the critical Kitaev's Majorana chain.  The bulk
interaction \eqnref{eq: Hint 2d} will induce the interaction
between Majorana surface modes, which corresponds to the coupling
of $\vect{n}^{(1)}$ and $\vect{n}^{(2)}$ at the boundary:
\begin{equation}
S'_\text{cp} = \int \dd\tau\dd x_2 A' \sum_{a=1,2,3} n_a^{(1)}n_a^{(2)} -
B' n_4^{(1)}n_4^{(2)}. \label{1dboundaryint}
\end{equation}
The $A'$ term corresponds to exactly the same fermion interaction
that trivialized eight copies of Majorana chain in the previous
section, and this coupling can gap out the critical point in the
previous $1d$ case. This means that the $A'$ term can also gap out
the boundary of the 8 copies of $2d$ $p \pm ip$ TSC without
degeneracy. Once the boundary is gapped and nondegenerate, a weak
$B'$ term in \eqnref{1dboundaryint} will not close the gap of the
boundary.
Since the boundary coupling \eqnref{1dboundaryint} is induced by
the bulk interaction \eqnref{eq: Hint 2d}, this implies that the
interaction in \eqnref{eq: Hint 2d} (with strong enough strength)
can gap out the bulk criticality (with 16 copies of $2d$ massless
Majorana fermions) in $2d$.

\subsection{From $^3$He-B to $3d$ Bosonic SPT}

\subsubsection{Lattice Model and Bulk Theory}

Let us go one dimension higher, and consider the $^3$He superfluid
B phase\cite{He3Bbalian,He3Bleggett1,He3Bleggett2,He3Bleggett3}
(will be denoted as $^3$He-B) which is a $3d$ TSC protected by the
$\dsZ_2^T$ symmetry with $\mathcal{T}^2=-1$ (symmetry class
DIII)\cite{AZclass}. The $^3$He-B TSC is described by the
following Hamiltonian
\begin{equation}\label{eq: 3d TSC}
H=\sum_{\vect{k}}\Big(\frac{\vect{k}^2}{2m_\text{He}}-\mu\Big)c_{\vect{k}}^\dagger
c_{\vect{k}}-\frac{\Delta}{2}(c_{-\vect{k}}\ii\sigma^2\vect{k}\cdot\vect{\sigma}c_{\vect{k}}+h.c.),
\end{equation}
where
$c_{\vect{k}}=(c_{\vect{k}\uparrow},c_{\vect{k}\downarrow})^\intercal$
is the fermion operator for the $^3$He atom, and $\Delta\in\dsR$
is the $p$-wave pairing strength. The Hamiltonian is invariant
under the time-reversal $\dsZ_2^T$ symmetry, which acts as
$\mathcal{T}: c_{\vect{k}}\to \ii\sigma^2 c_\vect{-k}$ followed by
the complex conjugation. $^3$He-B TSC corresponds to the $\mu>0$
topological phase of the model, while for $\mu<0$ the model
describes a trivial superconductor. Switching to the Majorana
basis
$\chi_{\vect{k}}=(c_{\vect{k}\uparrow}+c_{-\vect{k}\uparrow}^\dagger,-c_{\vect{k}\downarrow}-c_{-\vect{k}\downarrow}^\dagger,\ii
c_{\vect{k}\uparrow}-\ii c_{-\vect{k}\uparrow}^\dagger,-\ii
c_{\vect{k}\downarrow}+\ii
c_{-\vect{k}\downarrow}^\dagger)^\intercal/\sqrt{2}$ and in the
long-wave-length limit (to the first order in $\vect{k}$), the
effective Hamiltonian reads
\begin{equation}
H_{\times1}=\frac{1}{2}\sum_\vect{k}\chi_\vect{k}^\intercal(-k_1\sigma^{33}-k_2\sigma^{10}-k_3\sigma^{31}+m\sigma^{20})\chi_\vect{k},
\end{equation}
where we have set $\Delta=1$ as our energy unit, and defined the
topological mass $m=-\mu$ (which should not be confused with the
mass of the $^3$He atom $m_\text{He}$). The time-reversal operator
acting on the Majorana basis is given by
$\mathcal{T}=\mathcal{K}\ii\sigma^{32}$. The trivial ($m>0$) and
the topological ($m<0$) phases are separated by the phase
transition at $m=0$ where the bulk gap closes. This bulk
criticality is protected by the $\dsZ_2^T$ symmetry.

The $^3$He-B TSC belongs to the $3d$ DIII class fFSPT states,
which is known to be $\dsZ$
classified,\cite{ludwigclass1,ludwigclass2} and are indexed by the
topological number\cite{qi2008,wang1}
\begin{equation}
N=\int\frac{\dd^3\vect{k}}{8\pi^2}\Tr\sigma^{32}G\partial_{k_1}G^{-1}G\partial_{k_2}G^{-1}G\partial_{k_3}G^{-1},
\end{equation}
where $G(k)=-\langle\chi_\vect{k}\chi_\vect{-k}^\intercal\rangle$
is the fermion Green's function at zero frequency $\ii\omega=0$.
The $^3$He-B TSC in \eqnref{eq: 3d TSC} corresponds to $N=1$.
While the other topological states in this $\dsZ$ classification
may be realized by considering multiple copies of the $^3$He-B
TSC's, which can be described by the following effective field
theory Hamiltonian
\begin{equation}\label{eq: 3d H fFSPT}
\begin{split}
H_{\times\nu}=\frac{1}{2}\sum_{\alpha=1}^{\nu}\int\dd^3 \vect{x}\chi_\alpha^\intercal(&\ii\partial_1\sigma^{33}+\ii\partial_2\sigma^{10}\\
&+\ii\partial_3\sigma^{31}+m\sigma^{20})\chi_\alpha,
\end{split}
\end{equation}
with the $\dsZ_2^T$ symmetry protection
($\mathcal{T}=\mathcal{K}\ii\sigma^{32}$). $\nu$-copy $^3$He-B TSC
would correspond to the topological number $N=\nu$.

\subsubsection{Corresponding BSPT state}

However with interaction, the classification of the $3d$ DIII
class FSPT states is reduced from $\dsZ$ to $\dsZ_{16}$, meaning
that sixteen copies of the $^3$He-B TSC ($N=16$) can be smoothly
connected to the trivial state ($N=0$) in the presence of
interaction. This interaction reduced classification was discussed
in Ref.\,\onlinecite{chenhe3B,senthilhe3}, but here we will
provide another argument for it by making connection to the $3d$
BSPT states.

Let us start by showing that eight copies of the $^3$He TSC
($\nu=8$) can be connected to the $3d$ BSPT state with $\dsZ_2^T$
symmetry. Similar to our previous approach in $1d$ and $2d$, here
we should introduce five fermion pairing terms and couple them to
an O(5) order parameter field $\vect{n}=(n_1,n_2,n_3,n_4,n_5)$,
the low-energy effective FSM Hamiltonian reads
\begin{equation}\label{eq: 3d TSC FSM}
\begin{split}
H_{\times8}=&\,\frac{1}{2}\int\dd^3 \vect{x}\;\chi^\intercal h_{\times8}\chi,\\
h_{\times8}=&\,\ii\partial_1\sigma^{33000}+\ii\partial_2\sigma^{10000}+\ii\partial_3\sigma^{31000}\\
&+m\sigma^{20000}+n_1\sigma^{32212}+n_2\sigma^{32220}\\
&+n_3\sigma^{32232}+n_4\sigma^{32300}+n_5\sigma^{32100},
\end{split}
\end{equation}
where $\chi=(\chi_1,\chi_2,\cdots,\chi_8)^\intercal$. It turns out
that these order parameters are spin-singlet $s$-wave (time-reversal broken)
imaginary pairing among the eight copies of fermions. The
particular form of the pairing terms given here is not a unique
choice. We only require that the pairing terms anti-commute with
each other, and also anti-commute with the momentum and the
topological mass terms. However any other set of such pairing
terms are related to the above choice by basis transformation
among the eight copies of fermions, so we may stick to our current
choice without losing any generality.

On this $\nu=8$ Majorana basis, the time-reversal operator is
extended to $\mathcal{T}=\mathcal{K}\ii\sigma^{32000}$, from
which, it is easy to see that all five $s$-wave pairing terms
change sign under $\mathcal{T}$. To preserve the $\dsZ_2^T$
symmetry, we must require the order parameters to change sign as
well, \emph{i.e.} $\vect{n}\to-\vect{n}$, under the $\dsZ_2^T$
transform. After integrating out the fermion field $\chi$, we
arrive at the effective theory for the boson field $\vect{n}$,
which is a NLSM with a topological $\Theta$ term at $\Theta=2\pi$,
given by the following action ($\dd^4x=\dd\tau\dd^3\vect{x}$)
\begin{equation}\label{eq: 3d NLSM}
\begin{split}
S[\vect{n}]=&\int \dd^4 x\;\frac{1}{g}(\partial_\mu
\vect{n})^2\\
&+\frac{\ii\Theta}{\Omega_4}\epsilon^{abcde}n_a\partial_0
n_b\partial_1n_c\partial_2n_d\partial_3n_e,
\end{split}
\end{equation}
where $\Omega_4=8\pi^2/3$ is the volume of $S^4$. \eqnref{eq: 3d
NLSM} describes\cite{xuclass} a non-trivial $3d$ BSPT state\cite{senthilashvin, xu3dspt, xubf} protected by the $\dsZ_2^T$ symmetry
$\vect{n}\to-\vect{n}$. Thus we have established a connection from
eight copies of the $^3$He-B TSC to a single copy of the $3d$
$\dsZ_2^T$ BSPT state, bridging the FSPT and BSPT states in $3d$.

Now we can discuss the iFSPT states using the knowledge about the
BSPT states. On the BSPT side, based on the well-known $\dsZ_2$
classification of this state~\cite{wenspt2,senthilashvin}, it is
expected that two copies of the $\dsZ_2^T$ BSPT state can be
smoothly connected to the trivial state without breaking the
symmetry. In our NLSM formalism, this conclusion can be drawn by
the following inter-layer coupling
\begin{equation}
\begin{split}
S=&S[\vect{n}^{(1)}]+S[\vect{n}^{(2)}]+S_\text{cp},\\
S_\text{cp}=&\int \dd^4 x\;A\sum_{a=1,2,3}n_a^{(1)}n_a^{(2)}-B
n_4^{(1)}n_4^{(2)} -C n_5^{(1)}n_5^{(2)}. \label{scp}
\end{split}
\end{equation}
It is easy to check that the coupling $S_\text{cp}$ respects the
$\dsZ_2^T$ symmetry. When $A,B,C\to+\infty$, $\vect{n}^{(1)}$ and
$\vect{n}^{(2)}$ are locked anti-ferromagnetically for their first
three components and ferromagnetically for their last two
components, \emph{i.e.} $n_{a}^{(1)} = -n_{a}^{(2)} = n_a$
($a=1,2,3$) and $n_{b}^{(1)} = n_{b}^{(2)} = n_b$ ($b=4,5$). Then
the effective NLSM for the combined field $\vect{n}$ has $\Theta =
0$ due to the cancelation of the $\Theta$ angles between the two
layers. So two copies of the $\dsZ_2^T$ BSPT state can be
trivialized by the $A, B, C\to +\infty$ coupling. Again, this is
the necessary condition for interaction to reduce the
classification for $^3$He-B phase to $\dsZ_{16}$.

\subsubsection{Bulk Phase Transition under Interaction}

Now we would like to argue that the quantum critical point in the
noninteracting limit can be gapped out by interaction for 16
copies of $^3$He-B states. We start from the critical point $m=0$
in the FSM \eqnref{eq: 3d TSC FSM}, where the bulk gap is closed
on the free fermion level. The field theory \eqnref{eq: 3d TSC
FSM} at $m = 0$ has an extra inversion symmetry $\mathcal{P} = -\mathcal{I}\ii
\sigma^{32000}$ (where the space inversion operator $\mathcal{I}$ sends $\vect{x}\to-\vect{x}$), besides the
original time-reversal symmetry $\mathcal{T} = \mathcal{K} \ii
\sigma^{32000}$. Fermion interactions will be generated after
integrating out dynamical field $\vect{n}$. We will argue that in
this particular field theory \eqnref{eq: 3d TSC FSM}, interaction
can gap out the critical point, without driving the system into
either $m < 0$ or $m > 0$ state.

We can first gap out the fermions in the bulk by setting up a
fixed configuration of the order parameter field $\vect{n}$ at the
cost of breaking the time-reversal symmetry. Then we restore the
symmetry by proliferating the topological defects of the
$\vect{n}$ field, which is an approach adopted by
Ref.~\onlinecite{senthilhe3,xuSM}. Here we consider the
point defect, namely the monopole configuration of $\vect{n}$, which is described by $n_{a}\sim x_{a}$ (for $a=1,2, 3$)
and $n_{4}=n_{5}=0$ near the monopole core. This monopole breaks
both $\mathcal{T}$ and $\mathcal{P}$, but it preserves the combined symmetry
$\mathcal{T}' = \mathcal{P} \mathcal{T}$. After
proliferating this monopole, all the symmetries will be restored.

However the potential obstacle is that the monopole may trap
Majorana zero modes and is therefore degenerated. Proliferating
such defect will not result in a gapped and non-degenerated ground
state, and hence fails to gap out the bulk criticality. So we must
analyze the fermion modes at the monopole core carefully. By
solving the BdG equation for a single copy of the FSM \eqnref{eq:
3d TSC FSM}, it can be shown that the monopole will trap four
Majorana zero modes, which transforms under $\mathcal{T}^\prime$
as $\mathcal{T}^\prime: \gamma_a \rightarrow \gamma_a$, with $a =
1, \cdots 4$, followed by complex conjugation and space inversion. Thus for two copies of FSM \eqnref{eq: 3d TSC FSM},
the monopole will trap eight Majorana zero modes, and the
$\mathcal{T}^\prime$ symmetry will guarantee the spectrum of the
monopole is degenerate at the noninteracting level. Nevertheless
the degeneracy can be completely lifted by interaction\cite{fidkowski1,fidkowski2} without
breaking $\mathcal{T}^\prime$. So
after the monopole proliferation, all the symmetries of
\eqnref{eq: 3d TSC FSM} are restored, and the system will enter a
fully gapped state which still resides on the line $m=0$.
Therefore with two copies of the FSM, the iFSPT state can be
smoothly connected to the trivial state via strong interaction,
resulting in the $\dsZ_{16}$ classification, which is consistent
with the NLSM analysis.

Later we will show that this analysis of bulk phase transition
using topological defects can be naturally generalized to all
higher dimensions.

\subsubsection{Boundary Modes and Bulk transition}

Similar to what has been discussed in the $1d$ and $2d$ cases, the
inter-layer coupling $S_\text{cp}$ in \eqnref{scp} can be
immediately ported to the fermion model as a four-fermion local
interaction (with $A,B,C>0$)
\begin{equation}\label{eq: Hint 3d}
H_\text{int}=\int\dd^3\vect{x}\;A\sum_{a=1,2,3}\Delta_{a}^{(1)}\Delta_{a}^{(2)}
- B\Delta_{4}^{(1)}\Delta_{4}^{(2)} -
C\Delta_{5}^{(1)}\Delta_{5}^{(2)},
\end{equation}
where $\vect{\Delta}=\chi^\intercal
(\sigma^{32212},\sigma^{32220},\sigma^{32232},\sigma^{32300},\sigma^{32100})
\chi$ are defined for both layers of the $\nu=8$ fermions. We
should expect that sixteen copies of the $^3$He-B TSC can be
connected to the trivial state under this interaction, as the same
interaction can trivialize the BSPT in the NLSM.

Following the same idea of the $2d$ case, we argue that the
interaction can remove the $3d$ bulk criticality by showing that
its $2d$ boundary states can be symmetrically gapped out under
interaction.

Let us consider a boundary of the $3d$ system along the
$x_1$-$x_3$ plane, \emph{i.e.} the topological mass $m\sim x_2$
changes sign across $x_2=0$. For eight copies of the $^3$He-B TSC
as described in \eqnref{eq: 3d TSC FSM}, the boundary states are
given by the projection operator
$P=(1-\sigma^{10000}\sigma^{20000})/2$, such that the effective
FSM Hamiltonian along the boundary is given by
\begin{equation}
\begin{split}
H'_{\times8}=&\,\frac{1}{2}\int\dd x_1\dd x_3 \eta^\intercal h'_{\times 8}\eta,\\
h'_{\times8}=&\,\ii\partial_1\sigma^{3000}+\ii\partial_3\sigma^{1000}+n_1\sigma^{2212}+n_2\sigma^{2220}\\
&+n_3\sigma^{2232}+n_4\sigma^{2300}+n_5\sigma^{2100},
\end{split}
\end{equation}
where $\eta$ denotes the Majorana surface modes, and the
$\dsZ_2^T$ symmetry act as
$\mathcal{T}=\mathcal{K}\ii\sigma^{2000}$ on $\eta$. Under a
series of basis transformation as follows
\begin{equation}
\begin{split}
\eta\to & e^{\frac{\ii\pi}{4}\sigma^{1302}}
e^{-\frac{\ii\pi}{4}\sigma^{1002}}
e^{\frac{\ii\pi}{4}\sigma^{2300}}
e^{-\frac{\ii\pi}{4}\sigma^{2000}}\\&
e^{\frac{\ii\pi}{4}\sigma^{3102}}
e^{-\frac{\ii\pi}{4}\sigma^{0302}} \eta,
\end{split}
\end{equation}
the boundary Hamiltonian can be reformulated as
\begin{equation}\label{eq: 3d TSCx8 boundary}
\begin{split}
h'_{\times8}=&\,\ii\partial_1\sigma^{3000}+\ii\partial_3\sigma^{1300}+n_1\sigma^{1112}+n_2\sigma^{1120}\\
&+n_3\sigma^{1132}+n_4\sigma^{1200}+n_5\sigma^{2000},
\end{split}
\end{equation}
which, at the field theory level, is equivalent to four copies of
$2d$ (critical) $p\pm\ii p$ TSC described by \eqnref{eq: 2d TSC
FSM}, with the transformed $\dsZ_2^T$ symmetry
$\mathcal{T}=\mathcal{K}\ii\sigma^{2300}$. $(n_1,n_2,n_3,n_4)$ is
the analogue of the O(4) order parameter of the $p\pm\ii p$ TSC.
So once again, the problem is reduced by one dimension, to the
critical $2d$ iFSPT states.

If we consider the boundary of sixteen copies of the $^3$He-B TSC,
it will simply be a doubling of \eqnref{eq: 3d TSCx8 boundary},
which is analogous to eight copies of (critical) $p\pm\ii p$ TSC
at the field theory level. The bulk interaction \eqnref{eq: Hint
3d} will induce the interaction between Majorana surface
modes, which corresponds to the coupling of $\vect{n}^{(1)}$ and
$\vect{n}^{(2)}$ at the boundary:
\begin{equation}
\begin{split}
S'_\text{cp}=\int\dd\tau\dd x_1\dd x_3
\;&A'\sum_{a=1,2,3}n_{a}^{(1)}n_{a}^{(2)}\\
& -
B'n_{4}^{(1)}n_{4}^{(2)} -
C'n_{5}^{(1)}n_{5}^{(2)}.
\end{split}
\end{equation}
As we already argued, the $A'$ and $B'$ term together can gap out
the quantum critical point for 8 copies of $2d$ $p\pm ip$ TSC,
this means that the same interaction in the field theory can also
gap out the $2d$ boundary of 16 copies of $3d$ $^3$He-B phase. And
after the boundary is gapped out, adding a $C'$ term will not close
the gap at the boundary. So there should be no obstacle to
smoothly connect sixteen copies of $^3$He-B TSC to the trivial
state, under the interaction that is ported from inter-layer
coupling for the corresponding $3d$ $\dsZ_2^T$ BSPT states, $i.e.$
the $3d$ bulk interaction in \eqnref{eq: Hint 3d} can gap out the
SPT to trivial state quantum critical point of 16 copies of
$^3$He-B.

As one can see clearly now, the same pattern of logic will appear
again and again in every spatial dimension. Using the dimension
reduction argument, the boundary of $d$-dimensional iFSPT state
can be viewed, at the field theory level, as the
$(d-1)$-dimensional (critical) iFSPT. If the $(d-1)$-dimensional
criticality can be gapped out by a $(d-1)$-dimensional
interaction, then the same kind of interaction will be able to
trivialize the boundary of the $d-$dimensional iFSPT state, and
the $d-$dimensional bulk interaction that induces this
$(d-1)$-dimensional boundary interaction likely gaps out the
$d-$dimensional bulk criticality. Of course, one should be
reminded that we are not saying that the $d$-dimensional iFSPT
boundary has a $(d-1)$-dimensional lattice realization, our
induction is only based on the effective field theory description
of the long-wave-length physics. Following this induction
approach, a class of the iFSPT states and their interaction
reduced classification can be studied systematically in all
dimensions.


\section{SPT States with $\dsZ_2^P$ Only}

\subsection{General Constructions}

\subsubsection{Boson SPT with $\dsZ_2^P$}

Boson SPT (BSPT) state with inversion symmetry exists in all
dimensions. The construction is based on the O($d+2$) non-linear
$\sigma$ model (NLSM) in $(d+1)$-dimensional space-time with a
topological $\Theta$-term at $\Theta=2\pi$
\begin{equation}\label{eq: S BSPT Z2P}
\begin{split}
S[\vect{n}]=&\int \dd^{d+1} x \frac{1}{g}(\partial_\mu
\vect{n})^2+\frac{\ii\Theta}{\Omega_{d+1}}\epsilon^{a_1a_2a_3\cdots a_{d+2}}\\
&n_{a_1}\partial_{0}n_{a_2}\partial_{1}n_{a_3}\cdots\partial_{d}n_{a_{d+2}},
\end{split}
\end{equation}
where $x_0\equiv \tau$ is the time coordinate and the rest of $x_i$'s
($i=1,\cdots,d$) are space coordinates, and
$\Omega_{d+1}=2\pi^{\frac{d+2}{2}}/\Gamma(\frac{d+2}{2})$ is the
volume of a $(d+1)$-hypersphere with unit radius. The action of
the inversion symmetry $\dsZ_2^P$ inverts the space and flips all
components of $\vect{n}$,
\begin{equation}\label{eq: Z2P act NLSM}
\mathcal{P}:\left\{\begin{array}{ll} x_i\to-x_i & \text{for
}i=1,\cdots,d \\ n_a\to-n_a & \text{for
}a=1,\cdots,d+2\end{array}\right..
\end{equation}
It is straight forward to check that the action \eqnref{eq: S BSPT
Z2P} is invariant under this inversion. In the $g\to\infty$
regime, the model has a unique gapped disordered ground state,
which is a non-trivial SPT state when
$\Theta=2\pi$.\cite{xu2dspt,xu3dspt,xuclass}

This BSPT state is $\dsZ_2$ classified, meaning that two copies of
such state can be smoothly connected to the trivial state without
breaking the symmetry. To show this, we first make two copies of
the model in \eqnref{eq: S BSPT Z2P}, with $\vect{n}$-vectors
denoted by $\vect{n}^{(1)}$ and $\vect{n}^{(2)}$ in each copy
respectively, such that the total action reads
$S=S[\vect{n}^{(1)}]+S[\vect{n}^{(2)}]$. Then we are allowed to
turn on the following inversion symmetric coupling between the two
copies,
\begin{equation}\label{eq: A term}
S_\text{cp}=\int\dd^{d+1}x\;  A
n_{1}^{(1)}n_{1}^{(2)} - B\sum_{a=2}^{d+2}
n_{a}^{(1)}n_{a}^{(2)},
\end{equation}
In the limit of $A,B\to+\infty$, $\vect{n}^{(1)}$ and
$\vect{n}^{(2)}$ are locked together, and the final theory has
effectively $\Theta = 0$, and it is a trivial state. Thus the BSPT
with inversion symmetry is classified by $\dsZ_2$ within the
framework of NLSM.

However in two dimensional space (perhaps in some higher
dimensions as well), there are additional $\dsZ$ classified BSPT
states beyond NLSM, such as the $E_8$ state in 2d.
So the classifications in $d=2 \mod 4$ dimensions should be
extended to $\dsZ_2 \times \dsZ$.

\subsubsection{Free Fermion SPT with $\dsZ_2^P$}

Free fermion SPT (fFSPT) state with inversion symmetry also exists
in all dimensions, described by the quadratic Majorana Hamiltonian
\begin{equation}\label{eq: H fFSPT Z2P}
\begin{split}
H=&\frac{1}{2}\int\dd^d \vect{x}\; \chi^\intercal h_{\times1}\chi,\\
h_{\times1}=&\sum_{i=1}^{d}\ii\partial_i\alpha^i+m\beta^0 ,
\end{split}
\end{equation}
where $\chi$ denotes the Majorana fermion operator. $\alpha^i$ are
symmetric matrices while $\beta^0$ is anti-symmetric, and they all
anti-commute with each other. The action of the inversion symmetry
is given by the operator $\mathcal{P}=\mathcal{I}\ii\beta^0$,
where $\mathcal{I}$ is the space inversion operator such that
$\mathcal{I}^{-1}x_i\mathcal{I}=-x_i$ for all $i=1,\cdots,d$, and
it is followed by an orthogonal transform $\ii\beta^0$ in the
Majorana basis, where $\beta^0$ is just the mass matrix. Note that
this inversion symmetry acts as $\mathcal{P}^2=-1$ on the Majorana
fermions.

The $\dsZ_2^P$ fFSPT state belongs to the symmetry class D, and is
$\dsZ$ classified in general (see Table II in
Ref.\,\onlinecite{luZ2P}). Because $\mathcal{P}=\mathcal{I}\ii
\beta^0$ rules out all the other additional mass terms that
anti-commute with the topological mass $m\beta^0$, so one has to
go through a bulk phase transition (by closing the single-particle
gap) to drive the SPT state trivial (\emph{i.e.} to change the
sign of $m$). The exception rests in $d=2\mod 4$ dimensions, where
the classification is extended to $\dsZ\times\dsZ$, which was
pointed out in Ref.\,\onlinecite{luZ2P}, and will be discussed in
more details later.

Although the field theory Hamiltonian in \eqnref{eq: H fFSPT Z2P}
only describes the low-energy physics, it can be immediately cast
into lattice models by the substitution $\ii \partial_i\to \sin
k_i$ and $m\to\sum_{i=1}^d\cos k_i-d+m$, with $k_i$ being the
quasi-momentum of the fermion on the lattice. Some lattice models
has been explicitly constructed in Ref.\,\onlinecite{luZ2P}.

\subsubsection{Interacting Fermion SPT with $\dsZ_2^P$}

The interacting fermion SPT (iFSPT) states can be obtained by
introducing inversion symmetric interaction terms to the free
fermion Hamiltonian in \eqnref{eq: H fFSPT Z2P}. As we discussed
previously, interaction can reduce the classification of FSPT
states,
the same phenomenon is expected here. To study the interaction
reduced classification, we still make use of the BSPT states
discussed in the last section, and connect the iFSPT to BSPT by
introducing bosonic $\vect{n}$ degrees of freedoms:
\begin{equation}\label{eq: S iFSPT}
\begin{split}
S&=\int\dd^{d+1}x\; \frac{1}{2}\chi^{\intercal}(\ii\partial_0+h_{\times\nu})\chi+\frac{1}{g}(\partial_\mu \vect{n})^2+\cdots,\\
&h_{\times\nu}=\sum_{i=1}^{d}\ii\partial_i\alpha^i+m\beta^0
+\sum_{a=1}^{d+2}n_a\beta^a,
\end{split}
\end{equation}
where $h_{\times\nu}$ describes $\nu$ copies of the fFSPT in
\eqnref{eq: H fFSPT Z2P} coupling to the bosonic fields $n_a$. 
Here $\beta^a$ are anti-commuting anti-symmetric matrices, and
they also anti-commute with the all matrices $\alpha^i$ and
$\beta^0$ (which has been enlarged from those in \eqnref{eq: H
fFSPT Z2P} by tensor product with $\nu\times\nu$ identity matrix).
Of course, we need enough flavors of fermions (by making enough
$\nu$ copies of the fFSPT states) in order to support the $d+2$
additional $\beta^a$ matrices. Integrating out the boson field
$\vect{n}$, \eqnref{eq: S iFSPT} gives a pure fermionic model with
interaction. While integrating out the fermion field $\chi$,
\eqnref{eq: S iFSPT} becomes the NLSM as in \eqnref{eq: S BSPT
Z2P}. So \eqnref{eq: S iFSPT} establishes a connection between the
iFSPT and the BSPT phases~\cite{xubf}.

The inversion symmetry act as
\begin{equation}
\mathcal{P}:\left\{\begin{array}{ll} \vect{x}\to-\vect{x} \\
\chi\to\ii\beta^0\chi \\ \vect{n}\to-\vect{n}\end{array}\right..
\end{equation}
It is straight forward to verify that the action in \eqnref{eq: S
iFSPT} respects this inversion symmetry. The inversion symmetry
satisfies $\mathcal{P}^2=+1$ on the bosonic $\vect{n}$ vector, but
acts projectively as $\mathcal{P}^2=-1$ on the Majorana fermion.
With this set up, we can study the classification of iFSPT by
resorting to the classification of BSPT states, which are much
better understood.

\subsection{Examples in Each Dimension}
\subsubsection{$d=1$}

The $\dsZ_2^P$ fFSPT phase in $d=1$ is classified by $\dsZ$, and
its root state (Kitaev's Majorana chain) is described by the
lattice model \eqnref{eq: Maj chain real space} (at $\nu=1$). The
Hamiltonian is invariant under a bond centered inversion symmetry
$\dsZ_2^P$ (see \figref{fig: Majorana chain}), which acts on the
Majorana fermions as
$\mathcal{P}:\chi_{A}\to\chi_{B},\chi_{B}\to-\chi_{A}$. Because
the bond is directed due to the imaginary hopping $\ii u_{ij}$,
the inversion not only takes the fermion from $A$ sites to $B$
sites and vice versa, but must also be followed by a gauge
transformation, and hence we can have $\mathcal{P}^2=-1$ here. The
inversion operator can be also written as
$\mathcal{P}=\mathcal{I}\ii\sigma^2$ in the basis
$\chi=(\chi_A,\chi_B)^\intercal$ with
$\mathcal{I}^{-1}x\mathcal{I}=-x$ implements the inversion of the
spacial coordinate.

The low-energy effective Majorana Hamiltonian for the root state
is given by \eqnref{eq: 1d H fFSPT} (at $\nu=1$), and we repeat
here
\begin{equation}
h_{\times1}=\ii\partial_1\sigma^1+m\sigma^2,
\end{equation}
with $\mathcal{P}=\mathcal{I}\ii\sigma^2$. As long as this
inversion symmetry is preserved, without interaction, the two
sides of the phase diagram $m
> 0$ and $m < 0$ are always separated by a gapless critical point
at $m = 0$, no matter how many copies of the system we make.

To incorporate an O(3) order parameter, the model must be copied
four times
\begin{equation}
h_{\times4}=\ii\partial_1\sigma^{100}+m\sigma^{200}+n_1\sigma^{312}+n_2\sigma^{320}+n_3\sigma^{332},
\end{equation}
with $\mathcal{P}=\mathcal{I}\ii\sigma^{200}$ acting on the
Majorana basis. Under $\mathcal{P}$, we must also require
$\vect{n}\to -\vect{n}$. Then if we integrate out the fermions,
the effective theory becomes the O(3) NLSM at $\Theta=2\pi$
(Haldane spin chain). If we double the model $h_{\times4}$ again
to $h_{\times8}$, eight copies of this FSPT root state can be
trivialized by interaction (see section II for detailed analysis),
as the corresponding BSPT state is trivial due to its $\dsZ_2$
classification. Thus the $\dsZ_2$ classification of the BSPT
suggests that the $\dsZ_2^P$ iFSPT in $d=1$ is $\dsZ_8$
classified.

\subsubsection{$d=2$}
The $\dsZ_2^P$ fFSPT phase in $d=2$ is classified by
$\dsZ\times\dsZ$, which has two root states. They are $p+\ii p$
and $p-\ii p$ topological superconductors (TSC) respectively. The
free fermion $p+\ii p$ TSC in 2d is already $\dsZ$ classified
without any symmetry protection. However with the inversion
symmetry, the $p+\ii p$ and $p-\ii p$ TSC's are not allowed to
trivialize with each other, thus we will have two independent
$\dsZ$ topological indices $\nu_1$ and $\nu_2$ labelling the
copies of the $p+\ii p$ and $p-\ii p$ TSC's respectively.

To study the interaction reduced classification of this FSPT, we
start from the special case when $\nu_1=\nu_2$ (\emph{i.e.} the
non-chiral $p\pm\ii p$ TSC). The low-energy effective Majorana
Hamiltonian for the non-chiral root state is given by \eqnref{eq:
2d H fFSPT} (at $\nu=1$), and we repeat here
\begin{equation}
h_{\times1}=\ii\partial_1\sigma^{30}+\ii\partial_2\sigma^{13}+m\sigma^{20},
\end{equation}
with $\mathcal{P}=\mathcal{I}\ii\sigma^{20}$. In this case, the
iFSPT can be studied by making connection to the BSPT within the
scope of FSM. To incorporate an O(4) order parameter, the model
must be copied four times
\begin{equation}
\begin{split}
h_{\times4}=&\ii\partial_1\sigma^{3000}+\ii\partial_2\sigma^{1300}+m\sigma^{2000}\\
&+n_1\sigma^{1112}+n_2\sigma^{1120}+n_3\sigma^{1132}+n_4\sigma^{1200},
\end{split}
\end{equation}
with $\mathcal{P}=\mathcal{I}\ii\sigma^{2000}$ acting on the
Majorana basis. Under $\mathcal{P}$, we must also require
$\vect{n}\to -\vect{n}$. Then if we integrate out the fermions,
the effective theory becomes the O(4) NLSM at $\Theta=2\pi$. If we
double the model $h_{\times4}$ again to $h_{\times8}$, eight
copies of this FSPT root state can be trivialized by interaction (as
was discussed in section II), as the corresponding BSPT state is
trivial due to its $\dsZ_2$ classification.

When $\nu_1\neq\nu_2$, the iFSPT state cannot be connected to a
BSPT as discussed above, as no order parameter can be embedded no
matter how many copies of the fFSPT state we make. However we can
consider such FSPT state as attaching additional layers of chiral
$p+\ii p$ TSC (or $p-\ii p$ TSC) to the non-chiral $p\pm\ii p$
TSC's. It is known that interaction can not reduce the
classification of the $p+\ii p$ TSC, and the only possible effect
of the interaction is to drive 16 copies of the $p+\ii p$ TSC to a
BSPT state known as the $E_8$ state~\cite{xubf}. So we can simply
extend the $\dsZ_8$ classification of non-chiral iFSPT by
attaching the $\dsZ$ classified chiral iFSTP, and as a result, the
$\dsZ_2^P$ iFSPT in $d=2$ is $\dsZ_8\times\dsZ$ classified.
Correspondingly the BSPT in $d=2$ is $\dsZ_2\times\dsZ$, in which
the $\dsZ$ index labels the number of $E_8$ states.

\subsubsection{$d=3$}
The $\dsZ_2^P$ fFSPT phase in $d=3$ is classified by $\dsZ$. The
low-energy effective Majorana Hamiltonian for the root state is
given by \eqnref{eq: 3d H fFSPT} (at $\nu=1$), and we repeat here
\begin{equation}
h_{\times1}=\ii\partial_1\sigma^{33}+\ii\partial_2\sigma^{10}+\ii\partial_3\sigma^{31}+m\sigma^{20},
\end{equation}
with $\mathcal{P}=\mathcal{I}\ii\sigma^{20}$. To incorporate an
O(5) order parameter, the model must be copied eight times
\begin{equation}
\begin{split}
h_{\times8}=&\,\ii\partial_1\sigma^{33000}+\ii\partial_2\sigma^{10000}+\ii\partial_3\sigma^{31000}+m\sigma^{20000}\\
&+n_1\sigma^{32212}+n_2\sigma^{32220}+n_3\sigma^{32232}\\
&+n_4\sigma^{32300}+n_5\sigma^{32100},
\end{split}
\end{equation}
with $\mathcal{P}=\mathcal{I}\ii\sigma^{20000}$ acting on the
Majorana basis. Under $\mathcal{P}$, we must also require
$\vect{n}\to -\vect{n}$. Then if we integrate out the fermions,
the effective theory becomes the O(5) NLSM at $\Theta=2\pi$, which
is equivalent to the BSPT with $\dsZ_2^P$ symmetry. If we double
the model $h_{\times8}$ again to $h_{\times16}$, sixteen copies of
this FSPT root state can be trivialized by interaction, as the
corresponding BSPT state is trivial due to its $\dsZ_2$
classification. Therefore the $\dsZ_2^P$ iFSPT in $d=3$ is
$\dsZ_{16}$ classified.

\subsubsection{Higher Dimensions}

The above examples can be systematically generalized to higher
dimensions using the representation of Clifford algebras. In
$d$-dimensional space, the non-chiral root state of the $\dsZ_2^P$
fFSPT phase is described by the following Majorana Hamiltonian at
low-energy
\begin{equation}
h_{\times1}=\sum_{i=1}^{d}\ii\partial_i\alpha^i+m\beta^0,
\end{equation}
with $d$ symmetric matrices $\alpha^i$ and one anti-symmetric
matrix $\beta^0$, which are taken from the generators of the real
Clifford algebra $\Cl_{d,1}$, \emph{i.e.}
$\{\alpha^1,\cdots,\alpha^d;\ii\beta^0\}$ (see Appendix A for
definitions). The inversion symmetry acts as
$\mathcal{P}=\mathcal{I}\ii\beta^0$.

To construct the FSM in $d$-dimension, we must make enough copies
of the fFSPT root state to incorporate the  O($d+2$) order
parameter. Suppose $\nu$ is the minimal copies that should be
made, we can write down the following Majorana Hamiltonian
\begin{equation} \label{dFSM}
h_{\times\nu}=\sum_{i=1}^{d}\ii\partial_i\alpha^i+m\beta^0+\sum_{a=1}^{d+2}n_a\beta^a,
\end{equation}
in which the symmetric matrices $\alpha^i$ ($i=1,\cdots,d$) and
the anti-symmetric matrices $\beta^a$ ($a=1,\cdots,d+2$) can be
taken from the generators of the real Clifford algebra
$\Cl_{d,d+2}$, \emph{i.e.}
$\{\alpha^1,\cdots,\alpha^d;\ii\beta^1,\cdots,\ii\beta^{d+2}\}$
(see Appendix A for definitions), and the mass matrix $\beta^0$ is
chosen to be the pseudo scalar of $\Cl_{d,d+2}$, \emph{i.e.}
$\ii\beta^0=\prod_{i=1}^{d}\alpha^i\prod_{a=1}^{d+2}(\ii\beta^a)$.
It is straight forward to verify that the matrix $\beta^0$ is
anti-symmetric by definition of $\Cl_{d,d+2}$, and hence qualified
as a mass term. The inversion symmetry still acts as
$\mathcal{P}=\mathcal{I}\ii\beta^0$, and under $\mathcal{P}$, we
require $\vect{n}\to-\vect{n}$ as well, such that $h_{\times\nu}$
is inversion symmetric.

If we integrate out the fermions in the FSM $H=\int\dd^{d}x
\chi^\intercal h_{\times\nu}\chi$, the effective theory becomes a
NLSM of $\vect{n}$ at $\Theta=2\pi$. If we double $h_{\times\nu}$
again to $h_{\times2\nu}$, $2\nu$ copies of this $d$-dimensional
FSPT root state can be trivialized by interaction, as the corresponding
$d$-dimensional BSPT is also trivial due to its $\dsZ_2$
classification. So the non-chiral
$\dsZ_2^P$ iFSPT is classified by $\dsZ_{2\nu}$. The minimal copy
number $\nu$ will be determined in the following. However, we
recall that for $d=2\mod 4$, we also have the chiral FSPT states,
which fall outside the FSM-based classification.\cite{kapustin1} So the $\dsZ_2^P$
iFSPT states in $d=2\mod 4$ dimension will be
$\dsZ_{2\nu}\times\dsZ$ classified, where $\dsZ$ index labels the
chiral FSPT states.

\subsection{The Classification Table}

\subsubsection{Counting Minimal Copy Number}

The minimal copies $\nu$ that one should make to go from the iFSPT
root state to the FSM model can be simply determined from the
Majorana fermion flavor numbers of both models. Given that
$h_{\times1}$ and $h_{\times\nu}$ are constructed using the
irreducible representations of $\Cl_{d,1}$ and $\Cl_{d,d+2}$
respectively, the minimal copy number $\nu$ follows from
\begin{equation}
\nu=\frac{\dim \Cl_{d,d+2}}{\dim \Cl_{d,1}},
\end{equation}
where $\dim\Cl_{p,q}$ denotes the dimension of the irreducible
real representation of the real Clifford algebra $\Cl_{p,q}$. As
concluded in Appendix A, $\Cl_{d,d+2}\cong\dsH(2^d)$, so $\dim
\Cl_{d,d+2}=2^{d+2}$. Therefore we have $\nu=2^{d+2}/\dim
\Cl_{d,1}$, from which we can conclude the $\dsZ_{2\nu}$
classification of $\dsZ_2^P$ iFSPT states as in \tabref{tab: real
class}. The classification of iFSPT also shows the 8-fold Bott
periodicity.

\begin{table}[htdp]
\caption{The classification of $\dsZ_2^P$ iFSPT states in each
dimension $d$. The data of $\Cl_{d,1}$ (also see Appendix A) and
the minimal copy number $\nu$ are also listed.}
\begin{center}
\begin{tabular}{clll}
$d$ mod 8 & $\dim \Cl_{d,1}$ & $\nu$ & classification \\
\hline
0 & $\dim \dsC(2^{\frac{d}{2}})= 2^{\frac{d+2}{2}} $ & $2^\frac{d+2}{2}$ & $\dsZ_{2^\frac{d+4}{2}}$\\
1 & $\dim \dsR(2^{\frac{d+1}{2}})= 2^{\frac{d+1}{2}} $ & $2^\frac{d+3}{2}$ & $\dsZ_{2^\frac{d+5}{2}}$\\
2 & $\dim 2\dsR(2^{\frac{d}{2}})= 2^{\frac{d+2}{2}} $ & $2^\frac{d+2}{2}$ & $\dsZ_{2^\frac{d+4}{2}}\times\dsZ$\\
3 & $\dim \dsR(2^{\frac{d+1}{2}})= 2^{\frac{d+1}{2}} $ & $2^\frac{d+3}{2}$ & $\dsZ_{2^\frac{d+5}{2}}$\\
4 & $\dim \dsC(2^{\frac{d}{2}})= 2^{\frac{d+2}{2}} $ & $2^\frac{d+2}{2}$ & $\dsZ_{2^\frac{d+4}{2}}$\\
5 & $\dim \dsH(2^{\frac{d-1}{2}})= 2^{\frac{d+3}{2}} $ & $2^\frac{d+1}{2}$ & $\dsZ_{2^\frac{d+3}{2}}$\\
6 & $\dim 2\dsH(2^{\frac{d-2}{2}})= 2^{\frac{d+4}{2}} $ & $2^\frac{d}{2}$ & $\dsZ_{2^\frac{d+2}{2}}\times\dsZ$\\
7 & $\dim \dsH(2^{\frac{d-1}{2}})= 2^{\frac{d+3}{2}} $ &
$2^\frac{d+1}{2}$ & $\dsZ_{2^\frac{d+3}{2}}$
\end{tabular}
\end{center}
\label{tab: real class}
\end{table}

In $d\mod 4=2$ dimensions, the chiral FSPT states are not included
in this classifying
scheme~\cite{ludwigclass1,ludwigclass2,kitaevclass}. As the chiral
FSPT states can not be trivialize by the fermion interaction,
therefore they provide an additional $\dsZ$ classification. So we
conclude that the $\dsZ_2^P$ iFSPT states in $d\mod 4=2$
dimensions are $\dsZ_{2\nu}\times\dsZ$ classified.

\subsubsection{Bulk Phase Transition under Interaction}

Finally we would like to mention that the same classification for
the non-chiral states can be obtained by the same argument as in
section\,II.\,C.\,3. The idea is that if $2\nu$ copies of the FSPT root
state can be trivialized by fermion interaction, then there must
be a way to gap out its bulk phase transition with the trivial
FSPT state without breaking the symmetry.

Again we start from the critical point, which corresponds to $m=0$
in the FSM \eqnref{dFSM}. We first gap out the fermions in the
bulk by setting up a fixed configuration of the order parameter
field $\vect{n}$ at the cost of breaking the inversion symmetry.
Then we restore the symmetry by proliferating the
inversion-symmetric topological defects of the $\vect{n}$ field.
Here we choose to focus on the point defect, namely the monopole
configuration of $\vect{n}$, because such defect generally exists
in all dimensions. The monopole configuration is described by $n_{a}\sim
x_{a}$ (for $a=1,\cdots,d$) and $n_{d+1}=n_{d+2}=0$ near the
monopole core. Under inversion, both $\vect{n}$ and $\vect{x}$
changes sign, so the above monopole configuration is indeed
inversion-symmetric. Thus if we can proliferate such monopoles,
the inversion symmetry will be restored.

Again by solving the BdG equation for a single copy of the FSM, it
can be shown that the monopole will always trap four Majorana zero
modes no matter in which dimension $d$. This general property can
be simply verified by counting the fermion flavors. The
$d$-dimensional FSM has $\dim \Cl_{d,d+2}=2^{d+2}$ flavors of
Majorana fermions. Confining them to the core of a $d$-dimensional
monopole will reduce the fermion flavor number by $2^d$, so the
remaining flavor number is $2^{d+2}/2^d=4$. Thus for two copies of
FSM, the monopole will trap eight Majorana zero modes, whose
degeneracy is protected on the free fermion level by the inversion
symmetry left in the monopole core, {\it together} with the assumption
of $m=0$ at the critical point. Nevertheless the degeneracy can be
completely lifted by interaction,\cite{fidkowski1,fidkowski2} such
that the monopole can be trivialized. So after the monopole
proliferation, the inversion symmetry is restored, and we are left
with a gapped symmetric state at $m=0$. Therefore with two copies
of the FSM, the iFSPT state can be smoothly connected to the
trivial state via strong interaction, resulting in the
$\dsZ_{2\nu}$ classification, which is consistent with the NLSM
analysis.

\section{SPT States with $\dsZ_2^P$ Combined with Other Symmetries}

\subsection{$\U\times\dsZ_2^P$ SPT States}

\subsubsection{BSPT with $\U\times\dsZ_2^P$}

The $\U\times\dsZ_2^P$ BSPT states can be studied similarly as the
$\dsZ_2^P$ BSPT states under the framework of the O($d+2$) NLSM.
The inversion symmetry still flips all components of $\vect{n}$ as
in \eqnref{eq: Z2P act NLSM}. One remains to specify the $\U$
symmetry action as well. Based on our experiences from lower
dimensional cases~\cite{xuclass,senthilashvin}, the different ways
of imposing the $\U$ symmetry in the NLSM correspond to different
BSPT root states.

In odd dimension $d$, there are $(d+3)/2$ ways to impose the $\U$
symmetry transformation, labeled by $k=0,\cdots,(d+1)/2$ as
\begin{equation}\label{eq: U(1) NLSM}
\begin{split}
U_k:&(n_{2a-1}+\ii n_{2a})\to e^{\ii\theta}(n_{2a-1}+\ii n_{2a})\\
&\text{for }a=1,2,\cdots,k,
\end{split}
\end{equation}
whereas $k=0$ labels the case that the $\U$ symmetry has no action
on $\vect{n}$. Each assignment $U_k$ of the symmetry action leads
to a $\dsZ_2$ classification of the BSPT states, as two layers of
the BSPT root states can be trivialized via the inter-layer
coupling \eqnref{eq: A term} as argued previously. So the
$\U\times\dsZ_2^P$ BSPT states in odd dimension is
$\dsZ_2^{(d+3)/2}$ classified.

In even dimension $d$, there are $(d+4)/2$ ways to impose the $\U$
symmetry transformation, labeled by $k=0,\cdots,(d+2)/2$ following
the same definition in \eqnref{eq: U(1) NLSM}. For the first
$(d+2)/2$ implementations $U_{k}$ ($k=1,\cdots,d/2$), each leads
to a $\dsZ_2$ classification respectively. However the last
implementation $U_{(d+2)/2}$ leads to a $\dsZ$ classification, as
the coupling term \eqnref{eq: A term} would necessary breaks the
$U(1)$ symmetry, and is therefore forbidden, so that there is no
way to reduce the $\dsZ$ classification. Thus the
$\U\times\dsZ_2^P$ BSPT states in even dimension is
$\dsZ_2^{(d+2)/2}\times\dsZ$ classified.

We therefore conclude the classification of $\U\times\dsZ_2^P$
BSPT states in \tabref{tab: U(1) BSPT}. However, this
classification is not complete. The chiral states, such as $E_8$
states in $2d$, are not covered by the NLSM classification.
\begin{table}[htdp]
\caption{The classification of $\U\times\dsZ_2^P$ BSPT States}
\begin{center}
\begin{tabular}{cl}
$d$ mod 2 & classification\\
\hline
0 & $\dsZ_2^{(d+2)/2}\times\dsZ$\\
1 & $\dsZ_2^{(d+3)/2}$
\end{tabular}
\end{center}
\label{tab: U(1) BSPT}
\end{table}

\subsubsection{Free Fermion SPT with $\U\times\dsZ_2^P$}
With the $\U$ symmetry, the Majorana fermions $\chi$ can be paired
up to Dirac fermions $\psi=\chi'+\ii\chi''$, such that $\psi\to
e^{\ii\theta}\psi$ under the action of $\U$. The non-chiral
$\U\times\dsZ_2^P$ fFSPT root state can be described by the
following Dirac Hamiltonian at low-energy
\begin{equation}\label{eq: H fFSPT U1Z2P}
\begin{split}
H&=\int\dd^d\vect{x}\;\psi^{\dagger}\tilde{h}_{\times1}\psi,\\
\tilde{h}_{\times1}&=\sum_{i=1}^d\ii\partial_i\gamma^i+m\gamma^{d+1},
\end{split}
\end{equation}
in which $\gamma^i$ ($i=1,\cdots,d+1$) are Hermitian complex
matrices which anti-commute with each other. They can be taken
from the generators of the complex Clifford algebra $\Cl_{d+1}$
(in the complex representation). The action of inversion symmetry
is given by the operator $\mathcal{P}=\mathcal{I}\ii\gamma^{d+1}$
(if $\mathcal{P}^2=-1$) or by
$\mathcal{P}=\mathcal{I}\gamma^{d+1}$ (if $\mathcal{P}^2=+1$). In
the presence of the $\U$ symmetry, there is no essential
difference between $\mathcal{P}^2=-1$ and $\mathcal{P}^2=+1$ (as
they only differed by a $\U$ rotation which is part of the
symmetry), so we will focus on the former case.

The $\U\times\dsZ_2^P$ fFSPT state belongs to the symmetry class
A, and is $\dsZ$ classified in odd dimensions and $\dsZ\times\dsZ$
classified in even dimensions (see Table I, II in
Ref.\,\onlinecite{luZ2P}). Because the $\U$ symmetry rules out any
fermion pairing terms, and within the fermion hopping terms, the
inversion symmetry $\mathcal{P}$ forbids all the other possible
mass terms that anti-commute with the topological mass
$m\gamma^{d+1}$, so one has to go through a bulk transition (by
closing the single-particle gap) to drive the SPT state trivial
(\emph{i.e.} to change the sign of $m$). This explains the $\dsZ$
classification in odd dimensions, and one of the $\dsZ$
classification in even dimensions. While the other $\dsZ$
classification in even dimensions comes from the chiral state,
whose root state is described by the Dirac Hamiltonian
\begin{equation}
\tilde{h}_{\times1}=\sum_{i=1}^{d}
\ii\partial_{i}\gamma^i+m'\gamma^\text{ch},
\end{equation}
where $\gamma^\text{ch}\equiv\prod_{i=1}^d\gamma^i$ is the chiral
matrix (which exists only for even $d$). The chiral mass
$m'\gamma^\text{ch}$ also preserves the $\U\times\dsZ_2^P$
symmetry. It is impossible to find any additional $\U$ preserving
mass term that anti-commute with the chiral mass, so the chiral
states lead to the other $\dsZ$ classification in even dimensions.

\subsubsection{Interacting Fermion SPT with $\U\times\dsZ_2^P$}
The non-chiral $\U\times\dsZ_2^P$ iFSPT states can be also studied
by extending the fFSPT model to the FSM. In each dimension, the
FSM is still given by \eqnref{eq: S iFSPT} in the Majorana fermion
basis, with the matrices $\alpha^i$ ($i=1,\cdots,d$) and
$\beta^{a}$ ($a=1,\cdots,d+2$) taken form the generators of the
real Clifford algebra $\Cl_{d,d+2}$, and the mass matrix $\beta^0$
chosen to be the pseudo scalar of $\Cl_{d,d+2}$. We must make
enough copies of the $\U\times\dsZ_2^P$ fFSPT root states
described in \eqnref{eq: H fFSPT U1Z2P} to obtain the FSM. To
count the number $\nu$ of copies correctly, we first rewrite the
Hamiltonian in \eqnref{eq: H fFSPT U1Z2P} in the Majorana basis,
which takes the form of \eqnref{eq: H fFSPT Z2P}, with the
matrices $\alpha^{i}$ ($i=1,\cdots,d$) and $\beta^0$ still taken
from the generators of the complex Clifford algebra $\Cl_{d+1}$
but using its real representation. Then the minimal copy number
$\nu$ is given by
\begin{equation}\label{eq: nu U1}
\nu=\frac{\dim \Cl_{d,d+2}}{\dim \Cl_{d+1}},
\end{equation}
where $\dim \Cl_{n}$ denotes the dimension of the irreducible real
representation of the complex Clifford algebra $\Cl_{n}$. Given
that $\dim \Cl_{d,d+2}=2^{d+2}$ (see Appendix A), $\nu$ in each
dimension $d$ can be calculated in \tabref{tab: complex class}. We
conclude that the non-chiral $\U\times\dsZ_2^P$ fFSPT root state
can be made into a FSM incorporating the O($d+2$) order parameters
by copying $\nu$ times, such that their corresponding iFSPT states
can be classified by making connection to the BSPT
classifications. However, the chiral fFSPT root state (which
appears in even dimensions) can not be connected to the FSM
without breaking the $\U$ symmetry, and should be classified
separately.

\begin{table}[htdp]
\caption{The classification of $\U\times\dsZ_2^P$ iFSPT states in
each dimension $d$. The data of $\Cl_{d+1}$ (also see Appendix A)
and the minimal copy number $\nu$ are also listed.}
\begin{center}
\begin{tabular}{clll}
$d$ mod 2 & $\dim \Cl_{d+1}$ & $\nu$ & classification \\
\hline
0 & $\dim 2\dsC(2^{\frac{d}{2}})= 2^{\frac{d+4}{2}} $ & $2^\frac{d}{2}$ & $\dsZ_{2^\frac{d+2}{2}}\times\dsZ$\\
1 & $\dim \dsC(2^{\frac{d+1}{2}})= 2^{\frac{d+3}{2}} $ &
$2^\frac{d+1}{2}$ & $\dsZ_{2^\frac{d+3}{2}}$
\end{tabular}
\end{center}
\label{tab: complex class}
\end{table}

Integrating out the fermions in the FSM, we arrive at the O($d+2$)
NLSM in \eqnref{eq: S BSPT Z2P} at $\Theta=2\pi$, with the
symmetry action inherited from the FSM, such that the inversion
symmetry flips all components of $\vect{n}$, while the $\U$
symmetry rotates two and only two components of $\vect{n}$, say
$(n_1+\ii n_2)\to e^{\ii\theta}(n_1+\ii n_2)$ (see Appendix B for
examples). The action of $\U$ here corresponds to the $U_1$
implementation as defined in \eqnref{eq: U(1) NLSM}, which gives a
$\dsZ_2$ classification of the BSPT states, meaning that two
copies of the FSM can be trivialized by the interaction which also
trivialize the corresponding BSPT states. As have been counted in
\eqnref{eq: nu U1}, each copy of the FSM corresponds to $\nu$
copies of the fFSPT root states, so the non-chiral
$\U\times\dsZ_2^P$ iFSPT states are $\dsZ_{2\nu}$ classified (see
\tabref{tab: complex class}). For the chiral fFSPT states, it is
not possible to extend them to the FSM  without breaking the $\U$
symmetry, thus their classification can not be reduced by the
interaction which is still $\dsZ$. In conclusion, the
$\U\times\dsZ_2^P$ iFSPT states are $\dsZ_{2^\frac{d+3}{2}}$
classified in odd $d$ dimensions, and are
$\dsZ_{2^\frac{d+2}{2}}\times\dsZ$ classified in even $d$
dimensions.

\subsection{$\dsZ_2^T \times\dsZ_2^P$ SPT States}

\subsubsection{Boson SPT with $\dsZ_2^T\times\dsZ_2^P$}

We study the $\dsZ_2^T\times\dsZ_2^P$ BSPT states under the
framework of the O($d+2$) NLSM. The inversion symmetry $\dsZ_2^P$
always flips all components of $\vect{n}$ as in \eqnref{eq: Z2P
act NLSM}, while the time-reversal symmetry $\dsZ_2^T$ must flips
odd number of $\vect{n}$ components to keep the $\Theta$-term
invariant. Based on our experiences gained from lower dimentional
cases~\cite{xuclass}, the different ways of imposing the
$\dsZ_2^T$ symmetry in the NLSM correspond to different of BSPT
root states.

In odd dimension $d$, there are $(d+3)/2$ ways to impose the
$\dsZ_2^T$ symmetry transformation, labeled by
$k=0,\cdots,(d+1)/2$ as
\begin{equation}\label{eq: Z2T NLSM}
\mathcal{T}_k: n_i\to -n_i \text{ for }i=1,\cdots,2k+1,
\end{equation}
which flips the first $(2k+1)$-components of $\vect{n}$, leaving
the rest of the components unchanged. Each assignment
$\mathcal{T}_k$ of the symmetry action leads to a $\dsZ_2$
classification of the BSPT states, as two layers of the BSPT root
states can be trivialized via the inter-layer coupling \eqnref{eq:
A term} as argued previously. So the $\dsZ_2^T\times\dsZ_2^P$ BSPT
states in odd dimension is $\dsZ_2^{(d+3)/2}$ classified.

In even dimension $d$, there are $(d+2)/2$ ways to impose the $\U$
symmetry, labeled by $k=0,\cdots,d/2$ following the same
definition in \eqnref{eq: Z2T NLSM}. Each assignment
$\mathcal{T}_k$ still leads to a $\dsZ_2$ classification. Thus the
$\dsZ_2^T\times\dsZ_2^P$ BSPT states in even dimension is
$\dsZ_2^{(d+2)/2}$ classified.

We therefore conclude the classification of
$\dsZ_2^T\times\dsZ_2^P$ BSPT states in \tabref{tab: Z2T BSPT}.
Again this in not a complete classification, as the analogues of
the $E_8$ states are not considered here.
\begin{table}[htdp]
\caption{The classification of $\dsZ_2^T\times\dsZ_2^P$ BSPT States}
\begin{center}
\begin{tabular}{cl}
$d$ mod 2 & classification\\
\hline
0 & $\dsZ_2^{(d+2)/2}$\\
1 & $\dsZ_2^{(d+3)/2}$
\end{tabular}
\end{center}
\label{tab: Z2T BSPT}
\end{table}

\subsubsection{Free Fermion SPT with $\dsZ_2^T\times\dsZ_2^P$}
According to Ref.\,\onlinecite{luZ2P}, there is no $\dsZ$
classified free fermion $\dsZ_2^T\times\dsZ_2^P$ SPT states if the
time-reversal $\mathcal{T}$ and the inversion $\mathcal{P}$
commute with each other. For our purpose to study the interaction
reduced classification of FSPT states, we wish to start with
$\dsZ$ classified fFSPT states. Therefore we consider a peculiar
setting where $\mathcal{T}$ and $\mathcal{P}$ do not commute, and
the symmetry group (acting on the fermions) is defined by
\begin{equation}\label{eq: Z2TZ2P group}
\mathcal{T}^2=\mathcal{P}^2=-1,\quad
\mathcal{T}\mathcal{P}\mathcal{T}\mathcal{P}=-1.
\end{equation}
This is a projective representation of the
$\dsZ_2^T\times\dsZ_2^P$ symmetry, which can be realized as a
projective symmetry group \cite{wen2002} if the fermions are
coupled to a $\dsZ_2$ gauge field (such as spinons in the $\dsZ_2$
spin-liquid).

In the presence of the time-reversal symmetry, the chiral SPT
states are ruled out. The $\dsZ_2^T\times\dsZ_2^P$ fFSPT root
state is non-chiral, and can be described by the following
Majorana Hamiltonian at low-energy
\begin{equation}\label{eq: H fFSPT Z2TZ2P}
\begin{split}
H&=\frac{1}{2}\int\dd^d\vect{x}\;\chi^{\intercal}h_{\times1}\chi,\\
h_{\times1}&=\sum_{i=1}^d\ii\partial_i\alpha^i+m\beta^{1},
\end{split}
\end{equation}
with the inversion symmetry $\mathcal{P}=\mathcal{I}\ii\beta^1$
and time-reversal symmetry $\mathcal{T}=\mathcal{K}\ii\beta^2$.
Here the symmetric matrices $\alpha^i$ ($i=1,\cdots,d$) and the
anti-symmetric matrices $\beta^1$, $\beta^2$ are taken from the
generators of the real Clifford algebra $\Cl_{d,2}$, \emph{i.e.}
$\{\alpha^1,\cdots,\alpha^d;\ii\beta^1,\ii\beta^2\}$ (see appendix
A for definitions). It worth mention that at dimensions $d=3, 7$
(mod 8), the representations $\Cl_{3,2}\cong\dsR(4)\oplus\dsR(4)$
and $\Cl_{7,2}\cong\dsH(8)\oplus\dsH(8)$ can be split into two
sub-algebras. Each sub-algebra is sufficient to faithfully
represent the anti-commutation relations among the generators. So
the minimal fermion flavor is only half of $\dim\Cl_{d,2}$ when
$d=3, 7$ (mod 8). For later convenient, we define the reduced
dimension $\rdim$ as the dimension of the minimal faithful representation
of the anti-commuting generators (but not the whole algebra),
which follows
\begin{equation}\label{eq: rdim}
\rdim\Cl_{p,q}\equiv\left\{\begin{array}{ll}
\frac{1}{2}\dim\Cl_{p,q} & p-q=1, 5 (\text{mod }8),\\
\dim\Cl_{p,q} & \text{otherwise.}
\end{array}\right.
\end{equation}
Thus in terms of the reduced dimension, the Majorana fermion
flavor number of the $\dsZ_2^T\times\dsZ_2^P$ root state in
\eqnref{eq: H fFSPT Z2TZ2P} is simply given by $\rdim\Cl_{d,2}$ in
dimension $d$.

Because the inversion symmetry $\mathcal{P}$ has ruled out all the
other possible mass terms that anti-commute with the topological
mass $m\beta^1$, which already leads to the $\dsZ$ classification,
and the additional time-reversal symmetry will not change the
classification. So the $\dsZ_2^T\times\dsZ_2^P$ symmetry defined
in \eqnref{eq: Z2TZ2P group} fFSPT states are $\dsZ$ classified.

\subsubsection{Interacting Fermion SPT with $\dsZ_2^T\times\dsZ_2^P$}
The (projective) $\dsZ_2^T\times\dsZ_2^P$ iFSPT states can be also
studied by extending the fFSPT model to the FSM. In each
dimension, the FSM is still given by \eqnref{eq: S iFSPT} in the
Majorana fermion basis, with the matrices $\alpha^i$
($i=1,\cdots,d$) and $\beta^{a}$ ($a=1,\cdots,d+2$) taken form the
generators of the real Clifford algebra $\Cl_{d,d+2}$, and the
mass matrix $\beta^0$ chosen to be the pseudo scalar of
$\Cl_{d,d+2}$. We must make enough copies of the
$\dsZ_2^T\times\dsZ_2^P$ fFSPT root states described in
\eqnref{eq: H fFSPT Z2TZ2P} to obtain the FSM. Then the minimal
copy number $\nu$ is given by
\begin{equation}\label{eq: nu Z2T}
\nu=\frac{\dim \Cl_{d,d+2}}{\rdim \Cl_{d,2}},
\end{equation}
where $\rdim \Cl_{d,2}$ is the reduced dimension defined in
\eqnref{eq: rdim}. Given that $\dim \Cl_{d,d+2}=2^{d+2}$ (see
Appendix A), $\nu$ in each dimension $d$ can be calculated in
\tabref{tab: DIII class}. We conclude that the
$\dsZ_2^T\times\dsZ_2^P$ fFSPT root state can be made into a FSM
incorporating the O($d+2$) order parameters by copying $\nu$
times, such that their corresponding iFSPT states can be
classified by making connection to the BSPT classifications.

\begin{table}[th]
\caption{The classification of $\dsZ_2^T\times\dsZ_2^P$ iFSPT
states in each dimension $d$. The data of $\Cl_{d,2}$ (also see
Appendix A) and the minimal copy number $\nu$ are also listed.}
\begin{center}
\begin{tabular}{clll}
$d$ mod 8 & $\rdim \Cl_{d,2}$ & $\nu$ & classification \\
\hline
0 & $\rdim \dsH(2^{\frac{d}{2}})= 2^{\frac{d+4}{2}} $ & $2^\frac{d}{2}$ & $\dsZ_{2^\frac{d+2}{2}}$\\
1 & $\rdim \dsC(2^{\frac{d+1}{2}})= 2^{\frac{d+3}{2}} $ & $2^\frac{d+1}{2}$ & $\dsZ_{2^\frac{d+3}{2}}$\\
2 & $\rdim \dsR(2^{\frac{d+2}{2}})= 2^{\frac{d+2}{2}} $ & $2^\frac{d+2}{2}$ & $\dsZ_{2^\frac{d+4}{2}}$\\
3 & $\rdim 2\dsR(2^{\frac{d+1}{2}})= 2^{\frac{d+1}{2}} $ & $2^\frac{d+3}{2}$ & $\dsZ_{2^\frac{d+5}{2}}$\\
4 & $\rdim \dsR(2^{\frac{d+2}{2}})= 2^{\frac{d+2}{2}} $ & $2^\frac{d+2}{2}$ & $\dsZ_{2^\frac{d+4}{2}}$\\
5 & $\rdim \dsC(2^{\frac{d+1}{2}})= 2^{\frac{d+3}{2}} $ & $2^\frac{d+1}{2}$ & $\dsZ_{2^\frac{d+3}{2}}$\\
6 & $\rdim \dsH(2^{\frac{d}{2}})= 2^{\frac{d+4}{2}} $ & $2^\frac{d}{2}$ & $\dsZ_{2^\frac{d+2}{2}}$\\
7 & $\rdim 2\dsH(2^{\frac{d-1}{2}})= 2^{\frac{d+3}{2}} $ &
$2^\frac{d+1}{2}$ & $\dsZ_{2^\frac{d+3}{2}}$
\end{tabular}
\end{center}
\label{tab: DIII class}
\end{table}

Integrating out the fermions in the FSM, we arrive at the O($d+2$)
NLSM in \eqnref{eq: S BSPT Z2P} at $\Theta=2\pi$, with the
symmetry action inherited from the FSM, such that the inversion
symmetry flips all components of $\vect{n}$, while the $\dsZ_2^T$
symmetry always flips odd number of $\vect{n}$ components (see
Appendix B for examples). Such an implementation of the symmetry
action always results in the $\dsZ_2$ classified the BSPT states,
meaning that two copies of the FSM can be trivialized by the
interaction which also trivialize the corresponding BSPT states.
As have been counted in \eqnref{eq: nu Z2T}, each copy of the FSM
corresponds to $\nu$ copies of the fFSPT root states, so the
$\dsZ_2^T\times\dsZ_2^P$ iFSPT states are $\dsZ_{2\nu}$ classified
(see \tabref{tab: DIII class}).

\section{Summary}

In this paper we systematically studied the classification of a large class of
strongly interacting fermionic and bosonic SPT states in all dimensions. And for all
the examples we considered in this paper, we argue that the
classification of BSPT states implies that short range
interactions can reduce the classification of FSPT states with the
same symmetry. Further more, using different methods, we argue
that certain interaction can gap out the critical point between
the FSPT state and trivial state in the noninteracting limit,
which implies that under interaction some FSPT states are driven
trivial, and it can be connected to the trivial state without bulk
phase transition.

\begin{acknowledgments}

We would like to acknowledge the helpful discussions with Yuan-Ming Lu, Alexei Y. Kitaev and Xiao-Gang Wen. The authors are supported by the the David and Lucile Packard
Foundation and NSF Grant No. DMR-1151208.

\end{acknowledgments}

\bibliography{inversion}
\bibliographystyle{apsrev}

\onecolumngrid
\appendix
\section{Irreducible Representation of Clifford Algebra}
The generators of the real Clifford algebra $\Cl_{p,q}$ can be
represented by a set of real matrices
$\{\alpha^1,\cdots,\alpha^p;\ii\beta^1,\cdots,\ii\beta^q\}$
anti-commuting with each other
\begin{equation}
\begin{array}{cc}
\alpha^i\alpha^j=-\alpha^j\alpha^i, \beta^i\beta^j=-\beta^j\beta^i & \text{for }i\neq j,\\
\alpha^i\beta^j=-\beta^j\alpha^i &\text{for any }i, j,
\end{array}
\end{equation}
among which the $\alpha^i$ matrices square to $1$ (as
$\alpha^i\alpha^i=1$), and the $\ii\beta^i$ matrices squares to
$-1$ (as $\ii \beta^i\ii\beta^i=-1$). We adopt this notation such
that both $\alpha^i$ and $\beta^j$ matrices are Hermitian, and can
be expressed as the direction product of Pauli matrices.
$\alpha^i$'s are real and transpose symmetric, and there are $p$
of them in the generators of $\Cl_{p,q}$; while $\beta^i$'s are
imaginary and transpose anti-symmetric, and there are $q$ of them
in the generators of $\Cl_{p,q}$.

The real Clifford algebras are isomorphic to the matrix algebras
of real numbers $\dsR$, complex numbers $\dsC$ or quaternions
$\dsH$. The first several examples include: $\Cl_{0,0}\cong\dsR$
whose irreducible representation is one-dimensional (just a real
number), $\Cl_{0,1}\cong\dsC$ generated by $\{\ii\sigma^2\}$
giving a two-dimensional irreducible representation ($2\times2$
real matrix), and $\Cl_{0,2}\cong\dsH$ generated by
$\{\ii\sigma^{12},\ii\sigma^{32}\}$ giving a four-dimensional
irreducible representation ($4\times4$ real matrix). Here
$\sigma^{ijk\cdots}=\sigma^i\otimes\sigma^j\otimes\sigma^k\otimes\cdots$
denotes the direct product of a series of Pauli matrices
$\sigma^0$, $\sigma^1$, $\sigma^2$ or $\sigma^3$. The irreducible
representations of the real Clifford algebra are concluded in
\tabref{tab: Clpq}, where $\dsR(N)$, $\dsC(N)$ and $\dsH(N)$
denote the algebras of $N\times N$ matrix over $\dsR$, $\dsC$ and
$\dsH$ respectively, and $2\dsR(N)$, $2\dsH(N)$ are shorthand
notations of $\dsR(N)\oplus\dsR(N)$, $\dsH(N)\oplus\dsH(N)$. The
larger Clifford algebra lying outside the table can be obtained by
the 8-fold Bott periodicity, namely
$\Cl_{p+8,q}\cong\Cl_{p,q+8}\cong\Cl_{p,q}\otimes \dsR(16)$. From
\tabref{tab: Clpq}, the dimension of the (real) irreducible
representation of the Clifford algebra can be easily read out as
$\dim \dsR(N)=N$, $\dim \dsC(N)=2N$, $\dim \dsH(N)=4N$, $\dim
2\dsR(N)=2N$, $\dim 2\dsH(N)=8N$.

\begin{table}[htdp]
\caption{Periodic Table of Real Clifford Algebras}
\begin{center}
\begin{tabular}{cc|cccccccc}
\multicolumn{2}{c|}{\multirow{2}{*}{$\Cl_{p,q}$}} & \multicolumn{8}{c}{$q$} \\
& & 0 & 1 & 2 & 3 & 4 & 5 & 6 & 7 \\
\hline \multirow{8}{*}{$p$}
& 0 & $\dsR$ & $\dsC$ & $\dsH$ & $2\dsH$ & $\dsH(2)$ & $\dsC(4)$ & $\dsR(8)$ & $2\dsR(8)$\\
& 1 & $2\dsR$ & $\dsR(2)$ & $\dsC(2)$ & $\dsH(2)$ & $2\dsH(2)$ & $\dsH(4)$ & $\dsC(8)$ & $\dsR(16)$\\
& 2 & $\dsR(2)$ & $2\dsR(2)$ & $\dsR(4)$ & $\dsC(4)$ & $\dsH(4)$ & $2\dsH(4)$ & $\dsH(8)$ & $\dsC(16)$ \\
& 3 & $\dsC(2)$ & $\dsR(4)$ & $2\dsR(4)$ & $\dsR(8)$ & $\dsC(8)$ & $\dsH(8)$ & $2\dsH(8)$ & $\dsH(16)$\\
& 4 & $\dsH(2)$ & $\dsC(4)$ & $\dsR(8)$ & $2\dsR(8)$ & $\dsR(16)$ & $\dsC(16)$ & $\dsH(16)$ & $2\dsH(16)$\\
& 5 & $2\dsH(2)$ & $\dsH(4)$ & $\dsC(8)$ & $\dsR(16)$ & $2\dsR(16)$ & $\dsR(32)$ & $\dsC(32)$ & $\dsH(32)$\\
& 6 & $\dsH(4)$ & $2\dsH(4)$ & $\dsH(8)$ & $\dsC(16)$ & $\dsR(32)$ & $2\dsR(32)$ & $\dsR(64)$ & $\dsC(64)$\\
& 7 & $\dsC(8)$ & $\dsH(8)$ & $2\dsH(8)$ & $\dsH(16)$ & $\dsC(32)$
& $\dsR(64)$ & $2\dsR(64)$ & $\dsR(128)$
\end{tabular}
\end{center}
\label{tab: Clpq}
\end{table}

The complex Clifford algebra $\Cl_{n}$ is much simpler, whose
generators can be represented by a set of complex matrices
$\{\gamma^1,\cdots,\gamma^n\}$, satisfying
\begin{equation}
\gamma^i\gamma^j=-\gamma^j\gamma^i (\text{for }i\neq j),
\gamma^i\gamma^i=1.
\end{equation}
The complex Clifford algebras are isomorphic to the matrix
algebras of complex numbers $\dsC$. For even $n$,
$\Cl_{2m}\cong\dsC(2^m)$; and for odd $n$,
$\Cl_{2m+1}\cong\dsC(2^m)\oplus\dsC(2^m)$ (or shorthanded as
$2\dsC(2^m)$). Their real irreducible representations are of the
dimensions: $\dim\dsC(2^m)=2^{m+1}$ and $\dim2\dsC(2^m)=2^{m+2}$.

\section{Fermion $\sigma$-Model in Each Dimension}
Here we enumerate the examples of fermion $\sigma$-model (FSM) in
each dimension with explicit matrix representation, and show how
the various symmetry action is embedded in the FSM. The action of
the FSM in $d$-dimension takes the following general form
\begin{equation}
\begin{split}
S&=\int\dd^{d+1}x\; \frac{1}{2}\chi^{\intercal}(\ii\partial_0+h^{(d)})\chi+\frac{1}{g}(\partial_\mu \vect{n})^2+\cdots,\\
h^{(d)}&=\sum_{i=1}^{d}\ii\partial_i\alpha^i+m\beta^0
+\sum_{a=1}^{d+2}n_a\beta^a,
\end{split}
\end{equation}
where $\alpha^i$ ($i=1,\cdots,d$) are transpose-symmetric
Hermitian matrices and $\beta^a$ ($a=0,\cdots,d+2$) are
transpose-antisymmetric Hermitian matrices. We consider that the
inversion symmetry always act as
$\mathcal{P}=\mathcal{I}\ii\beta^0$. We will provide the explicit
examples of these matrices in the fermion Hamiltonian $h^{(d)}$.
In general, the Majorana fermion $\chi$ is of $2^{d+2}$ flavors,
meaning that the dimensions of the matrices $\alpha^i$ and
$\beta^a$ are $2^{d+2}$. We will use the notation
$\sigma^{ijk\cdots}=\sigma^i\otimes\sigma^j\otimes\sigma^k\otimes\cdots$
to denote the direct product of Pauli matrices $\sigma^1$,
$\sigma^2$, $\sigma^3$ and as well as the $2\times2$ identity
matrix $\sigma^0$. Each Pauli matrix $\sigma^i$ acts on a
2-dimensional single-particle Hilbert space, which is also the
size of a qubit. We may thus count the dimension of the
single-particle Hilbert space (which is also the Majorana fermion
flavor number) by qubits. A Hamiltonian made of matrices like
$\sigma^{ijk\cdots}$  with $n$ indices acts in the Hilbert space
of $n$ qubit which is of the dimension $2^n$. In the following, we
will use examples to demonstrate both the FSM and the FSPT root
state model with various symmetries.


In $d=1$ spacial dimension, the Hamiltonian $h^{(1)}$ is defined
on a 3-qubit single-particle Hilbert space,
\begin{equation}
h^{(1)}=\ii\partial_1\sigma^{100}+m\sigma^{200}+n_1\sigma^{312}+n_2\sigma^{332}+n_3\sigma^{320}.
\end{equation}
With the $\dsZ_2^P: \chi\to\ii\sigma^{200}\chi,
\vect{n}\to-\vect{n}$ symmetry only, the root state Hamiltonian
takes the first 1-qubit subspace, and must be 4-multiplied to form
the FSM. With the additional
$\U:\chi\to\exp(\ii\varphi\sigma^{020})\chi,(n_1+\ii n_2)\to
e^{2\ii\varphi}(n_1+\ii n_2)$ symmetry, the root state Hamiltonian
takes the first 2-qubit subspace, and must be doubled to form the
FSM. With the additional
$\dsZ_2^T:\chi\to\ii\sigma^{320}\chi,n_3\to-n_3$ symmetry, the
root state Hamiltonian takes the first 2-qubit subspace, and must
be doubled to form the FSM.

In $d=2$ spacial dimension, the Hamiltonian $h^{(2)}$ is defined
on a 4-qubit single-particle Hilbert space,
\begin{equation}
h^{(2)}=\ii\partial_1\sigma^{1000}+\ii\partial_2\sigma^{3100}+m\sigma^{2000}+n_1\sigma^{3312}+n_2\sigma^{3332}+n_3\sigma^{3320}+n_4\sigma^{3200}.
\end{equation}
With the $\dsZ_2^P: \chi\to\ii\sigma^{2000}\chi,
\vect{n}\to-\vect{n}$ symmetry only, the root state Hamiltonian
takes the first 2-qubit subspace, and must be 4-multiplied to form
the FSM. With the additional
$\U:\chi\to\exp(\ii\varphi\sigma^{0020})\chi,(n_1+\ii n_2)\to
e^{2\ii\varphi}(n_1+\ii n_2)$ symmetry, the root state Hamiltonian
takes the first 3-qubit subspace, and must be doubled to form the
FSM. With the additional
$\dsZ_2^T:\chi\to\ii\sigma^{3200}\chi,n_4\to-n_4$ symmetry, the
root state Hamiltonian takes the first 2-qubit subspace, and must
be 4-multiplied to form the FSM.

In $d=3$ spacial dimension, the Hamiltonian $h^{(3)}$ is defined
on a 5-qubit single-particle Hilbert space,
\begin{equation}
h^{(3)}=\ii\partial_1\sigma^{10000}+\ii\partial_2\sigma^{31000}+\ii\partial_3\sigma^{33000}+m\sigma^{20000}+n_1\sigma^{32100}+n_2\sigma^{32300}+n_3\sigma^{32212}+n_4\sigma^{32232}+n_5\sigma^{32220}.
\end{equation}
With the $\dsZ_2^P: \chi\to\ii\sigma^{20000}\chi,
\vect{n}\to-\vect{n}$ symmetry only, the root state Hamiltonian
takes the first 2-qubit subspace, and must be 8-multiplied to form
the FSM. With the additional
$\U:\chi\to\exp(\ii\varphi\sigma^{00200})\chi,(n_1+\ii n_2)\to
e^{2\ii\varphi}(n_1+\ii n_2)$ symmetry, the root state Hamiltonian
takes the first 3-qubit subspace, and must be 4-multiplied to form
the FSM. With the additional
$\dsZ_2^T:\chi\to\ii\sigma^{32000}\chi,\vect{n}\to-\vect{n}$
symmetry, the root state Hamiltonian takes the first 2-qubit
subspace, and must be 8-multiplied to form the FSM.

In $d=4$ spacial dimension, the Hamiltonian $h^{(4)}$ is defined
on a 6-qubit single-particle Hilbert space,
\begin{equation}
\begin{split}
h^{(4)}&=\ii\partial_1\sigma^{100000}+\ii\partial_2\sigma^{310000}+\ii\partial_3\sigma^{330000}+\ii\partial_4\sigma^{322000}+m\sigma^{200000}\\&+n_1\sigma^{321100}+n_2\sigma^{321300}+n_3\sigma^{321212}+n_4\sigma^{321232}+n_5\sigma^{321220}+n_6\sigma^{323000}.
\end{split}
\end{equation}
With the $\dsZ_2^P: \chi\to\ii\sigma^{200000}\chi,
\vect{n}\to-\vect{n}$ symmetry only, the root state Hamiltonian
takes the first 3-qubit subspace, and must be 8-multiplied to form
the FSM. With the additional
$\U:\chi\to\exp(\ii\varphi\sigma^{000200})\chi,(n_1+\ii n_2)\to
e^{2\ii\varphi}(n_1+\ii n_2)$ symmetry, the root state Hamiltonian
takes the first 4-qubit subspace, and must be 4-multiplied to form
the FSM. With the additional
$\dsZ_2^T:\chi\to\ii\sigma^{323000}\chi,n_6\to-n_6$ symmetry, the
root state Hamiltonian takes the first 3-qubit subspace, and must
be 8-multiplied to form the FSM.

In $d=5$ spacial dimension, the Hamiltonian $h^{(5)}$ is defined
on a 7-qubit single-particle Hilbert space,
\begin{equation}
\begin{split}
h^{(5)}&=\ii\partial_1\sigma^{1000000}+\ii\partial_2\sigma^{3100000}+\ii\partial_3\sigma^{3300000}+\ii\partial_4\sigma^{3212000}+\ii\partial_5\sigma^{3232000}+m\sigma^{2000000}\\&+n_1\sigma^{3201000}+n_2\sigma^{3203000}+n_3\sigma^{3222100}+n_4\sigma^{322300}+n_5\sigma^{3222212}+n_6\sigma^{3222232}+n_7\sigma^{3222220}.
\end{split}
\end{equation}
With the $\dsZ_2^P: \chi\to\ii\sigma^{2000000}\chi,
\vect{n}\to-\vect{n}$ symmetry only, the root state Hamiltonian
takes the first 4-qubit subspace, and must be 8-multiplied to form
the FSM. With the additional
$\U:\chi\to\exp(\ii\varphi\sigma^{0002000})\chi,(n_1+\ii n_2)\to
e^{2\ii\varphi}(n_1+\ii n_2)$ symmetry, the root state Hamiltonian
takes the first 4-qubit subspace, and must be 8-multiplied to form
the FSM. With the additional
$\dsZ_2^T:\chi\to\ii\sigma^{3201000}\chi,n_1\to-n_1$ symmetry, the
root state Hamiltonian takes the first 4-qubit subspace, and must
be 8-multiplied to form the FSM.

In $d=6$ spacial dimension, the Hamiltonian $h^{(6)}$ is defined
on a 8-qubit single-particle Hilbert space,
\begin{equation}
\begin{split}
h^{(6)}&=\ii\partial_1\sigma^{10000000}+\ii\partial_2\sigma^{31000000}+\ii\partial_3\sigma^{33000000}+\ii\partial_4\sigma^{32120000}+\ii\partial_5\sigma^{32320000}+\ii\partial_6\sigma^{32222000}+m\sigma^{20000000}\\&+n_1\sigma^{32221000}+n_2\sigma^{32223000}+n_3\sigma^{32010100}+n_4\sigma^{32010300}+n_5\sigma^{32010212}+n_6\sigma^{32010232}\\
&+n_7\sigma^{32010220}+n_8\sigma^{32030000}.
\end{split}
\end{equation}
With the $\dsZ_2^P: \chi\to\ii\sigma^{20000000}\chi,
\vect{n}\to-\vect{n}$ symmetry only, the root state Hamiltonian
takes the first 5-qubit subspace, and must be 8-multiplied to form
the FSM. With the additional
$\U:\chi\to\exp(\ii\varphi\sigma^{00002000})\chi,(n_1+\ii n_2)\to
e^{2\ii\varphi}(n_1+\ii n_2)$ symmetry, the root state Hamiltonian
takes the first 5-qubit subspace, and must be 8-multiplied to form
the FSM. With the additional
$\dsZ_2^T:\chi\to\ii\sigma^{32030000}\chi,n_6\to-n_6$ symmetry,
the root state Hamiltonian takes the first 5-qubit subspace, and
must be 8-multiplied to form the FSM.

In $d=7$ spacial dimension, the Hamiltonian $h^{(7)}$ is defined
on a 9-qubit single-particle Hilbert space,
\begin{equation}
\begin{split}
h^{(7)}&=\ii\partial_1\sigma^{100000000}+\ii\partial_2\sigma^{310000000}+\ii\partial_3\sigma^{330000000}+\ii\partial_4\sigma^{321200000}+\ii\partial_5\sigma^{323200000}\\
&+\ii\partial_6\sigma^{320120000}+\ii\partial_7\sigma^{320320000}+m\sigma^{200000000}\\&+n_1\sigma^{322010002}+n_2\sigma^{322030002}+n_3\sigma^{322221200}+n_4\sigma^{322223200}+n_5\sigma^{322220120}+n_6\sigma^{322220320}\\
&+n_7\sigma^{322222010}+n_8\sigma^{322222030}+n_9\sigma^{322222220}.
\end{split}
\end{equation}
With the $\dsZ_2^P: \chi\to\ii\sigma^{200000000}\chi,
\vect{n}\to-\vect{n}$ symmetry only, the root state Hamiltonian
takes the first 5-qubit subspace, and must be 16-multiplied to
form the FSM. With the additional
$\U:\chi\to\exp(\ii\varphi\sigma^{000020000})\chi,(n_1+\ii n_2)\to
e^{2\ii\varphi}(n_1+\ii n_2)$ symmetry, the root state Hamiltonian
takes the first 5-qubit subspace, and must be 16-multiplied to
form the FSM. With the additional
$\dsZ_2^T:\chi\to\ii\sigma^{322200000}\chi,\vect{n}\to-\vect{n}$
symmetry, the root state Hamiltonian takes the first 5-qubit
subspace, and must be 16-multiplied to form the FSM.

In $d=8$ spacial dimension, the Hamiltonian $h^{(8)}$ is defined
on a 10-qubit single-particle Hilbert space,
\begin{equation}
\begin{split}
h^{(8)}&=\ii\partial_1\sigma^{1000000000}+\ii\partial_2\sigma^{3100000000}+\ii\partial_3\sigma^{3300000000}+\ii\partial_4\sigma^{3212000000}+\ii\partial_5\sigma^{3232000000}\\
&+\ii\partial_6\sigma^{3201200000}+\ii\partial_7\sigma^{3203200000}+\ii\partial_8\sigma^{3222200000}+m\sigma^{2000000000}\\&+n_1\sigma^{3220310002}+n_2\sigma^{3220330002}+n_3\sigma^{3220101200}+n_4\sigma^{3220103200}+n_5\sigma^{3220100120}+n_6\sigma^{3220100320}\\
&+n_7\sigma^{3220102010}+n_8\sigma^{3220102030}+n_9\sigma^{3220102220}+n_{10}\sigma^{3220320000}.
\end{split}
\end{equation}
With the $\dsZ_2^P: \chi\to\ii\sigma^{2000000000}\chi,
\vect{n}\to-\vect{n}$ symmetry only, the root state Hamiltonian
takes the first 5-qubit subspace, and must be 32-multiplied to
form the FSM. With the additional
$\U:\chi\to\exp(\ii\varphi\sigma^{0000020000})\chi,(n_1+\ii
n_2)\to e^{2\ii\varphi}(n_1+\ii n_2)$ symmetry, the root state
Hamiltonian takes the first 6-qubit subspace, and must be
16-multiplied to form the FSM. With the additional
$\dsZ_2^T:\chi\to\ii\sigma^{3220320000}\chi,n_{10}\to-n_{10}$
symmetry, the root state Hamiltonian takes the first 6-qubit
subspace, and must be 16-multiplied to form the FSM.

For higher spacial dimensions ($d>8$), due to the Bott periodicity
of the Clifford algebra, the Hamiltonian $h^{(d)}$ can be extended
systematically from the Hamiltonian $h^{(d-8)}$ with 8 dimensions
lower. Suppose
$h^{(d)}=\sum_{i=1}^{d}\ii\partial_i\alpha^i+m\beta^0+\sum_{a=1}^{d+2}n_a\beta^a$
and
$h^{(d-8)}=\sum_{i=1}^{d-8}\ii\partial_i\tilde{\alpha}^i+m\tilde{\beta}^0+\sum_{a=1}^{d-6}n_a\tilde{\beta}^a$,
the extension is given by
\begin{equation}
\begin{array}{c}
\alpha^i=\tilde{\alpha}^i\otimes\sigma^{00000000},\quad(i=1,\cdots,d-9)\\
\beta^a=\tilde{\beta}^a\otimes\sigma^{00000000},\quad(a=0,\cdots,d-7)\\
\begin{array}{rr}
\alpha^{d-8}=\tilde{\alpha}^{d-8}\otimes\sigma^{22220000}, &\beta^{d-6}=\tilde{\beta}^{d-6}\otimes\sigma^{00002222},\\
\alpha^{d-7}=\tilde{\alpha}^{d-8}\otimes\sigma^{10000000}, &\beta^{d-5}=\tilde{\beta}^{d-6}\otimes\sigma^{00001000},\\
\alpha^{d-6}=\tilde{\alpha}^{d-8}\otimes\sigma^{30000000}, &\beta^{d-4}=\tilde{\beta}^{d-6}\otimes\sigma^{00003000},\\
\alpha^{d-5}=\tilde{\alpha}^{d-8}\otimes\sigma^{21200000}, &\beta^{d-3}=\tilde{\beta}^{d-6}\otimes\sigma^{00002120},\\
\alpha^{d-4}=\tilde{\alpha}^{d-8}\otimes\sigma^{23200000}, &\beta^{d-2}=\tilde{\beta}^{d-6}\otimes\sigma^{00002320},\\
\alpha^{d-3}=\tilde{\alpha}^{d-8}\otimes\sigma^{20120000}, &\beta^{d-1}=\tilde{\beta}^{d-6}\otimes\sigma^{00002012},\\
\alpha^{d-2}=\tilde{\alpha}^{d-8}\otimes\sigma^{20320000}, &\beta^{d}=\tilde{\beta}^{d-6}\otimes\sigma^{00002032},\\
\alpha^{d-1}=\tilde{\alpha}^{d-8}\otimes\sigma^{22010000}, &\beta^{d+1}=\tilde{\beta}^{d-6}\otimes\sigma^{00002201},\\
\alpha^{d}=\tilde{\alpha}^{d-8}\otimes\sigma^{22030000},
&\beta^{d+2}=\tilde{\beta}^{d-6}\otimes\sigma^{00002203}.
\end{array}
\end{array}
\end{equation}
Every symmetry transform matrix $O$ is extended to
$O\otimes\sigma^{00000000}$. The structure of the Hamiltonian and
the symmetry action remains the same under the extension. The
single-particle Hilbert space of every root state is enlarged by 4
qubits, while that of the FSM is enlarged by 8 qubits, so the
minimal copy number to obtain the FSM from the root state is
always 16 times multiplied in every 8 dimensions higher.

\end{document}